\PassOptionsToPackage{utf8}{inputenc}
\documentclass{bioinfo}
\copyrightyear{2024} \pubyear{2024}

\access{Advance Access Publication Date: Day Month Year}
\appnotes{Manuscript Category}

\usepackage[colorlinks=true, citecolor=blue, linkcolor=blue, urlcolor=BrickRed]{hyperref}
\usepackage{amsmath,amssymb}
\usepackage{amsthm}
\usepackage{multicol}
\usepackage{colortbl}
\usepackage[dvipsnames]{xcolor}
\usepackage{algorithm}
\usepackage{algpseudocode}
\usepackage{tikz}
\usepackage{soul}
\usetikzlibrary{positioning}

\newcommand{\BP}{\text{BP}}
\newcommand{\RF}{\text{RF}}
\newcommand{\IL}{\text{IL}}
\newcommand{\TE}{\text{TE}}
\newcommand{\RE}{\text{RE}}
\newcommand{\DL}{\text{DL}}

\definecolor{softgreen}{HTML}{82e0aa}
\definecolor{darkgreen}{HTML}{145a32}
\definecolor{softred}{HTML}{f1948a}

\theoremstyle{definition}
\newtheorem{definition}{Definition}
\newtheorem{problem}{Problem}

\theoremstyle{proposition}
\newtheorem{theorem}{Theorem}
\newtheorem{lemma}{Lemma}
\newtheorem{proposition}{Proposition}

\begin{document}
\firstpage{1}

\subtitle{Research article}

\title[Median  and Small Parsimony Problems on RNA trees]{Median  and Small Parsimony Problems on RNA trees}
\author[Marchand \textit{et~al}.]{ Bertrand Marchand\,$^{\text{\sfb 1,}}$,Yoann Anselmetti\,$^{\text{\sfb 1,}}$, Manuel Lafond\,$^{\text{\sfb 1,}}$  and A\"ida Ouangraoua\,$^{\text{\sfb 1,}*}$}
\address{$^{\text{\sf 1}}$Department of Computer Science, University of Sherbrooke, 2500 Boulevard de l’Universit\'e,
Sherbrooke, QC J1K 2R1, Canada.}

\corresp{$^\ast$To whom correspondence should be addressed.}
%\corresp{$^\dag$Contributed equally.}
\history{Received on XXXXX; revised on XXXXX; accepted on XXXXX}

\editor{Associate Editor: XXXXXXX}

\abstract{
\textbf{Motivation:}
Non-coding RNAs (ncRNAs) express their functions by adopting molecular structures. Specifically, RNA secondary structures serve as a relatively stable intermediate step before tertiary structures, offering a reliable signature of molecular function. Consequently, within an RNA functional family, secondary structures are generally more evolutionarily conserved than sequences. Conversely, homologous RNA families grouped within an RNA clan share ancestors but typically exhibit structural differences. Inferring the evolution of RNA structures within RNA families and clans is crucial for gaining insights into functional adaptations over time and providing clues about the Ancient RNA World Hypothesis.\\
\textbf{Results:}
We introduce the median problem and the small parsimony problem for ncRNA families, where secondary structures are represented as leaf-labelled trees. We utilize the Robinson-Foulds (RF) tree distance, which corresponds to a specific edit distance between RNA trees, and a new metric called the Internal-Leafset (IL) distance. While the RF tree distance compares sets of leaves descending from internal nodes of two RNA trees, the IL distance compares the collection of leaf-children of internal nodes. The latter is better at capturing differences in structural elements of RNAs than the RF distance, which is more focused on base pairs. We also consider a more general tree edit distance that allows the mapping of base pairs that are not perfectly aligned.
We study the theoretical complexity of the median problem and the small parsimony problem under the three distance metrics and various biologically-relevant constraints, and we present polynomial-time maximum parsimony algorithms for solving some versions of the problems. Our algorithms are applied to ncRNA families from the RFAM database, illustrating their practical utility.\\
\textbf{Contact:} \href{aida.ouagraoua@usherbrooke.ca}{aida.ouagraoua@usherbrooke.ca}\\
\textbf{Availability and Implementation:
\url{https://github.com/bmarchand/rna\_small\_parsimony}
}
\textbf{Supplementary information:}
Supplementary data are available at \textit{Bioinformatics} online.
}

\maketitle

\section{Introduction}

The RNA World hypothesis proposes that RNA preceded DNA as the primary genetic material on Earth \citep{bartel1999constructing,higgs2015rna}. It suggests that RNA served both as a carrier of genetic information and as a catalyst for chemical reactions, playing a central role in the early development of life before the evolution of the current DNA-based genetic system. Over time, DNA emerged as a more complex and stable molecule, taking over as the primary genetic material, while proteins became the primary catalysts for biochemical reactions.

An RNA molecule is mathematically defined as a string over an alphabet of four letters $\{A,C,G,U\}$ corresponding to nucleotide bases. An RNA structure is defined by an RNA sequence and a set of canonical base pairs that form hydrogen bonds, resulting in the folding of the RNA sequence. The structure of RNA molecules is closely linked to their function \citep{mattick2005functional}. Studying the evolution of RNA structures aims to infer ancestral RNA sequences and structures to understand how these structures have adapted and evolved over time, leading to the diversity of RNA observed today \citep{gruber2008strategies,tremblay2016reconstruction,Holmes_2004}. It can provide information about the RNA World hypothesis and the roles played by RNA in early biological processes by revealing patterns of conservation and providing insights into functional adaptations of RNA structures. RNA structures also carry information about the evolutionary history of organisms. Understanding the evolution of RNA structures over time aids in reconstructing the evolutionary history of organisms.

Extensive work has been done on the reconstruction of ancestral sequences, which, given a phylogeny with observed sequences at its leaves, aims to find the sequences at the internal nodes and the changes that happened on the branches of the phylogeny, leading to the diversity of observed sequences \citep{Fitch_1971,Sankoff_1975,altschul1989trees,schultz1996reconstruction,fredslund2003large,blanchette2008computational}. To infer ancestral RNA sequences and structures, only two approaches have been proposed to date: In 2009, in a pioneering work, \cite{bradley2009evolutionary} developed a maximum likelihood approach to compute the ancestral RNA sequences and structures under the TKF Structure Tree model introduced in \cite{Holmes_2004}. The TKF Structure Tree model is a continuous-time model of RNA structure evolution that extends the Thorne, Kishino, and Felsenstein's 1991 (TKF91) model of sequence evolution \cite{thorne1991evolutionary}. It allows not only single nucleotide substitutions and indels but also base-pair substitutions and indels, as well as whole sub-structure indels. However, the computational burden involved in the inference algorithm makes it prohibitive in practice. In 2016, \cite{tremblay2016reconstruction} developed a maximum parsimony method to compute the ancestral RNA sequences under a structurally constrained single nucleotide substitution model, given an alignment of input sequences. The underlying evolution model allows only single nucleotide substitutions, and the method utilizes secondary structures as constraints. It considers a single consensus secondary structure for each RNA family and does not account for the structural differences between members of an RNA family.

\begin{figure}
    \centering
    \includegraphics[width=.6\linewidth]{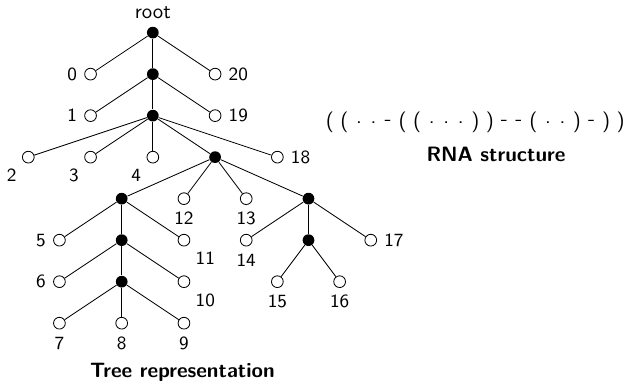}
    \caption{(Top) An RNA structure given in dot-bracket notation,
    and (Bottom) the corresponding tree representation, as defined formally in Definition~\ref{def:rna_tree}.}
    \label{fig:tree-representation}
\end{figure}

In this article, we introduce the median and small parsimony problems on RNA trees. An RNA secondary structure is an RNA structure such that the RNA segments defined by any two pairs of bases forming hydrogen bonds are either strictly nested or disjoint. It is a hierarchical structure composed of paired and unpaired bases at the fine-grain level, defining structural elements such as loops and stems at the coarse-grain level. RNA secondary structures can be represented in the form of ordered rooted trees. There exist several different tree representations for RNA secondary structures depending on the level of representation, fine- or coarse-grain, and also depending on the representation of base pairs as single or multiple nodes in the tree representation \citep{le1989tree,shapiro1990comparing,hochsmann2003local,ouangraoua2007local}. In this paper, we consider an RNA tree representation that extends the representation as a forest defined in \cite{hochsmann2003local}. Leaf nodes correspond to bases in the 5' to 3' primary sequence order, and internal nodes correspond to bonds between paired bases. Each internal node has at least two leaf-children, the leftmost and rightmost children, which correspond to the paired bases linked by a chemical bond, plus potentially other leaf-children that form a loop in the secondary structure and other children that are internal nodes (see Figure \ref{fig:tree-representation} for illustration).

\begin{figure*}
    \centering
    \includegraphics[width=\textwidth]{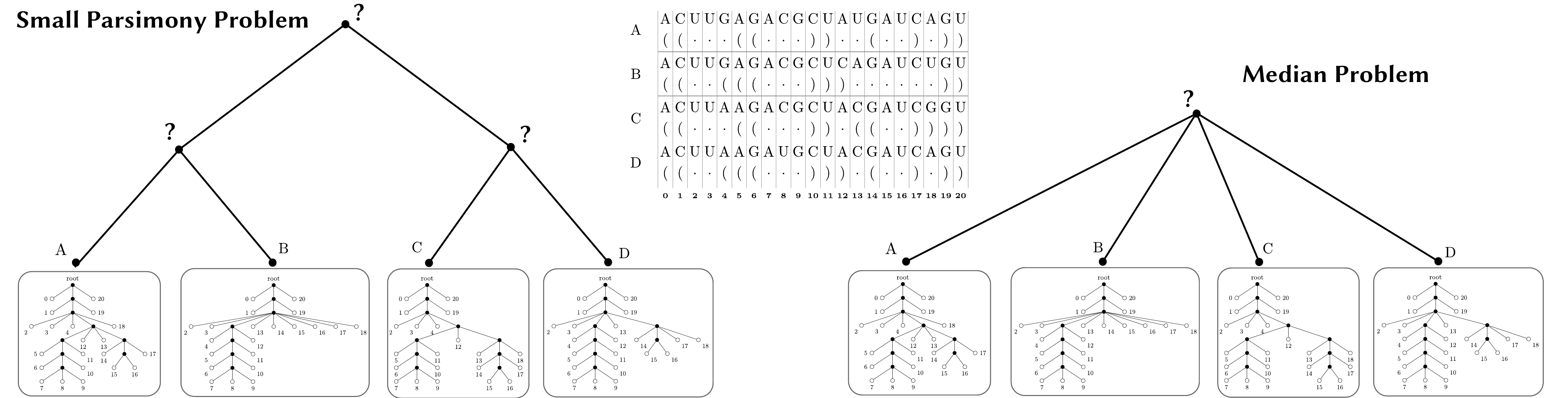}
    \caption{
    (Left) The Small Parsimony Problem consists in finding the best assignment of RNA structures to internal nodes of a given phylogeny, minimizing the sum of distances over edges. (Right) The Median Problem consists in finding a single RNA structure that minimizes the sum of distances to the input structures. 
    }
\vspace{-.5pt}
\label{fig:small_parsimony_median_illustration}

\end{figure*}

We investigate two problems. Given a multiple RNA secondary structure alignment, the Median Problem involves finding an RNA secondary structure that minimizes the sum of distances to the input RNA structures. On the other hand, the Small Parsimony Problem, given a multiple RNA secondary structure alignment and a phylogeny on the input RNA structures, aims to determine an optimal assignment of RNA secondary structures at the internal nodes of the phylogeny, minimizing the sum of distances on the branches of the phylogeny. The Median and Small Parsimony problems on biological sequences are often considered together because they are related in the context of phylogenetics and ancestral sequence reconstruction. Many heuristic algorithms designed for the Small Parsimony problem involve starting with initial assignments of ancestral sequences at the internal nodes. The algorithms then iteratively improve the assignment at one node at a time by considering its neighboring nodes and solving median subproblems \citep{sankoff1976frequency}. The definition of both problems is illustrated in Figure~\ref{fig:small_parsimony_median_illustration}.

Here, we consider three distance metrics between RNA secondary structures, given a multiple RNA secondary structure alignment. We focus on the restricted case of gap-free alignments, equivalent to considering a set of RNA structures with the same sequence length. In the case of a gapped multiple RNA structure alignment, a gap-free alignment can be obtained by removing all gapped columns from the alignment. If only one side of a base pair in an RNA secondary structure is removed, the remaining side is marked as an unpaired base. The resulting RNA structures will then have the same sequence length.

The first metric we consider is the base pair distance, defined as the size of the symmetric difference between the sets of base pairs of two RNA structures. In a model of RNA structure evolution, the base pair distance can also be defined as the minimum number of base pair breaking or creation events required to transform one RNA structure into the other. We demonstrate that the base pair distance between two RNA trees equals the \emph{Robinson-Foulds (RF) distance} between the corresponding RNA trees. The RF distance is a widely used metric for quantifying the dissimilarity between the topological structures of two phylogenetic trees \citep{robinson1981comparison}. Between two rooted trees on the same leaf set, it is defined as the size of the symmetric difference between the two collections of sets of leaves descending from internal nodes of the two trees. The RF distance between phylogenetic trees does not have a direct biological interpretation in terms of specific genetic evolution, but between RNA trees, it does have a meaning in terms of RNA structure evolution in the case of gap-free multiple RNA structure alignments. It corresponds to a scenario of internal node deletion and insertion operations between two RNA trees, representing a scenario of base pair breaking and creation events between two RNA structures.

The RF distance essentially accounts for the difference in terms of the presence or absence of base pairings between two RNA structures. Yet, the difference in terms of loops is also relevant when comparing two RNA structures. Loops constitute the structural units at the basis of the stability of RNA secondary structures, as illustrated by the Turner Energy Model (\cite{turner2010nndb}), considered the gold standard of RNA secondary structure energy models. Within this model, the free energy of a structure is a sum of contributions over individual loops, depending on their precise nucleotide compositions. In this sense, the presence or absence of a loop directly contributes to the stability of the molecule. Inspired by the RF distance, we introduce a second distance metric called the \emph{Internal-Leafset (IL) distance}, allowing us to account for the difference in terms of the presence or absence of loops. It is defined as the size of the symmetric difference between the two collections of leaf-children of internal nodes of the two trees.

The base pair distance between RNA secondary structures is known to be very strict \citep{schirmer2014introduction}. We show that it can be expressed as the Tree edit distance between the corresponding RNA trees under a specific cost scheme. The tree edit distance between two trees is the minimum cost of a mapping between the nodes of the two trees. The cost scheme under which the tree edit distance equals the base pair distance allows the mapping between two base pair nodes of two RNA trees only if the base pairs are perfectly aligned in the multiple RNA structure alignment. By relaxing the cost scheme to allow the mapping between base pairs that are not perfectly aligned, we consider a third distance metric called the \emph{Relaxed Edit (RE) distance}.

First, we present, in Section \ref{algo_median_pb}, polynomial-time algorithms for computing the median of RNA trees under the RF, IL, and RE distances and various biologically-relevant constraints on the output RNA tree. We then present, in Section \ref{algo_small_parsimony_pb}, polynomial-time exact and heuristic solutions for the small parsimony problem. Finally, in Section \ref{exp_results}, we illustrate the application of our approaches on simulated RNA families and RNA families from the RFAM database \cite{Rfam10v1, kalvari2021rfam}. For space reasons, some of the proofs and figures are given in Appendix.
%\ao{[Compléter avec résultats détaillés.]}

\vspace{-.5cm}
\section{Preliminaries}

We first introduce the definitions and notations required in the next sections.

\subsection{RNA Tree Representation}

An RNA secondary structure is represented by a pair $R=(S,P)$ where $S$ is a string of length $n$ on the alphabet $\Sigma = \{A, C, G, U\}$, and $P$ is a set of pairs $(i,j)$ such that $1 \leq i < j \leq n$. Any two pairs $(i,j)$ and $(k,l)$ in $P$ satisfy one of the following conditions: $i < k < l < j$, $k < i < j < l$, $i < j < k < l$, or $k < l < i < j$. We use the notation $[i, j] = \{i, i+1, \ldots, j\}$, which is the empty set if $i > j$.

We denote $L(R) = [0, n+1]$ as the set of base positions $[1, n]$ in $S$ augmented with two fictive start and end positions, and $I(R)$ as the set of pairs $P$ augmented with a fictive pair $(0, n+1)$, i.e., $I(R) = P \cup \{(0, n+1)\}$. The union of these two sets is denoted by $V(R) = L(R) \cup I(R)$. The nesting of base pairs defines a partial order relation on $V(R)$ denoted by $\prec$. For two pairs $(i,j)$ and $(k,l)$ in $I(R)$, $(i,j) \prec (k,l)$ if and only if $i < k < l < j$. For a pair $(i,j)\in I(R)$ and a base position $k\in L(R)$, $(i,j) \prec k$ if and only if $i \leq k \leq j$.

The tree representing an RNA secondary structure $R=(S,P)$ is an ordered rooted tree, denoted by $T(R)$, with its set of nodes being $V(R)$. The set of edges $E(R)$ is determined by the nesting of base pairs. For two nodes $x$ and $y$ in $V(R)$, there exists an edge $(x,y)\in E(R)$ if and only if $x \prec y$, and there is no third node $z \in V(R)$ such that $x \prec z \prec y$. Thus, $I(R)$ constitutes the internal nodes of the tree, with $(0,n+1)$ being the root, and $L(R)$ constitutes the leaf nodes. Moreover, for any internal node $(i,j) \in I(R)$, the set of leaf nodes $k \in L(R)$ descending from $(i,j)$, i.e., such that $(i,j) \prec k$, is $[i,j]$. The children of a node $(i,j) \in I(R)$ are ordered according to the left-to-right ordering of the set of leaf nodes descending from them. Therefore, the leftmost and rightmost child of the node $(i,j)$ are respectively the leaf nodes $i$ and $j$ (see Figure \ref{fig:tree-representation} for illustration).

Note that the partial order relation $\prec$ defined on $V(R)$ is the ancestor-descendant relation on the nodes of $T(R)$. For any two distinct nodes $x$ and $y$ in $V(R)$, $x \prec y$ if and only if $y$ is a child of $x$, or there exists a third node $z$ such that $x \prec z$ and $y$ is a child of $z$. In this case, $y$ is called a descendant of $x$. Given the ancestor-descendant relation $\prec$ defined on $V(R)$ and the total order relations defined on the set of children of each node $x$ in $V(R)$, we denote by $<$ the total order relation on $V(R)$ defined by a postorder traversal of the tree.

\begin{definition}[RNA tree]
    An RNA tree is an ordered rooted tree $T$ with an ordered leafset $L=[k,l]$, $0 \leq k < l$,  such that for any internal node $x$ of $T$, the leftmost and rightmost child of $x$ are two distinct leaves $i, j$, in which case $x = (i, j)$.

\label{def:rna_tree}
\end{definition}

Given an RNA tree $T$, $L(T)$ denotes its leafset, and $I(T)$ its set of internal nodes.  It is not difficult to see that RNA secondary structures and RNA trees are equivalent.

\begin{proposition}
Given an RNA secondary structure $R=(S,P)$ with $S$ of length $n$, the tree $T(R)$ is an RNA tree with leafset $[0,n+1]$. Given an RNA tree $T$ with leafset $L(T)=[0,n+1]$, $R= (S,P)$ where $S$ is a string of length $n$ on the alphabet $\Sigma$ and $P=I(T)-\{(0,n+1)\}$ is an RNA secondary structure.
\end{proposition}

Note also that, given an RNA tree $T$, $BP(T)=I(T)$ denotes the set of base pairs induced by $T$.

\subsection{Distance Between RNA trees}

Here, we consider two RNA secondary structures $R_1=(S_1,P_1)$ and $R_2=(S_2,P_2)$ of the same length $n$ and their RNA tree representations $T_1=(V(R_1), E(R_1))$ and $T_2=(V(R_2), E(R_2))$. Note that the trees $T_1$ and $T_2$ have the same leafset $L(R_1)=L(R_2) = [0,n+1]$.

\begin{definition}[Base pair distance]
    The base pair (BP) distance  between $R_1$ and $R_2$ is the size of the symmetric difference between $P_1$ and $P_2$:
    $$d_{\BP}(R_1,R_2) = |P_1 \Delta P_2| = |I(R_1) \Delta I(R_2)|.$$
\end{definition}

Given a rooted tree $T$ and an internal node $x$ of $T$, the \emph{descendant leafset} of $x$, denoted by $\DL(x)$, is the set of leaves $y$ of $T$ such that $x \prec y$. Note that if $T$ represents a phylogeny, $\DL(x)$ is also called a \emph{clade}. The \emph{internal leafset} of $x$, denoted by $\IL(x)$, is the set of children of $x$ that are also leaves of $T$.
We denote by $\DL(T)$, the collection of descendant leafsets of all internal nodes of $T$, $\DL(T) = \{\DL(x) | x \in I(T)\}$, while $\IL(T)$ denotes the collection of internal leafsets of all internal nodes of $T$, $\IL(T) = \{\IL(x) | x \in I(T)\}$.

We consider the following distances  between RNA trees $T_1$ and $T_2$.

\begin{definition}[Robinson-Foulds distance]
    The Robinson-Foulds (RF) distance  between  $T_1$ and $T_2$ is the size of the symmetric difference between $\DL(T_1)$ and $\DL(T_2)$:
    $$d_{\RF}(T_1,T_2)  = |\DL(T_1) \Delta \DL(T_2)|.$$
\end{definition}

\begin{definition}[Internal-Leafset distance]
  The Internal-Leafset (IL) distance  between  $T_1$ and $T_2$ is the size of the symmetric difference between $\IL(T_1)$ and $\IL(T_2)$:
    $$d_{\IL}(T_1,T_2)  = |\IL(T_1) \Delta \IL(T_2)|.$$
    \label{def:il_distance}
\end{definition}

A \emph{valid mapping} $M$ between $T_1$ and $T_2$ is a partial bijection between their sets of internal nodes $I(R_1)$ and $I(R_2)$ that preserves the order and the nesting relations of nodes. In other words, for any two pairs of mapped nodes $(x_1,x_2)$ and $(y_1,y_2)$ in $M$, we have $x_1 < y_1$ if and only if $x_2 < y_2$, and $x_1 \prec y_1$ if and only if $x_2 \prec y_2$. We denote by $\mathcal{M}(T_1,T_2)$ the set of all valid mappings between $T_1$ and $T_2$. A cost function $c : I(R_1) \times I(R_2) \rightarrow \mathbb{R}^+$ defines the cost of any couple in $I(R_1) \times I(R_2)$, such that $c(x_1,x_2) = 0$ if $x_1$ and $x_2$ represent two base pairs that are perfectly aligned, i.e., $x_1=x_2 = (i,j)$; otherwise, $c(x_1,x_2) > 0$.
Given a cost function $c$, the cost of a valid mapping $M$ between $T_1$ and $T_2$ is then defined as: $$cost_c(M) = \sum_{(x_1,x_2) \in M}{c(x_1,x_2)} + |I(R_1)| + |I(R_2)| - 2\times |M|.$$

\begin{definition}[Tree Edit distance]
    Given a cost function $c : I(R_1) \times I(R_2) \rightarrow \mathbb{R}^+$, the Tree Edit (TE) distance between $T_1$ and $T_2$  under the cost function $c$ is the minimum cost of a valid mapping between $T_1$ and $T_2$:
    $$d_{\TE_c}(T_1,T_2)  = \min_{M \in \mathcal{M}(T_1,T_2)}{cost_c(M)}.$$
\end{definition}

\begin{proposition}[Equality of BP distance and RF distance]
For any two RNA trees $T_1, T_2$: 
$$d_{\BP}(R_1,R_2) = d_{\RF}(T_1,T_2).$$
\end{proposition}

The above holds because
%, for a base pair $(i, j) \in I(R_1)$, the corresponding node $(i, j)$ of $T_1$ satisfies $\DL((i, j)) = [i,j]$. 
%There is therefore 
in an RNA tree, a base pair $(i, j)$ corresponds to the DL $[i, j]$.  Another relationship is the following.

\begin{lemma}[Equality of BP distance and TE distance under a specific cost function]
If the cost function $c$ is defined as $c((i_1,j_1),(i_2,j_2)) = 0$ if $(i_1,j_1)=(i_2,j_2)$, otherwise $c((i_1,j_1),(i_2,j_2)) = +\infty$, then
$d_{\BP}(R_1,R_2) = d_{\TE_c}(T_1,T_2)$.
\end{lemma}

\begin{proof}
The cost function heavily penalizes the mapping of different nodes of $T_1$ and $T_2$, i.e different base pairs of $R_1$ and $R_2$. Therefore, an optimal mapping cannot contain a pair of mapped nodes that are different. The optimal mapping minimizing $cost_c(M)$ is then $M=\{((i,j),(i,j)) ~|~ (i,j) \in I(R_1)\cap I(R_2))\}$, and thus
$cost_c(M) = |I(R_1)| + |I(R_2)| - 2\times | I(R_1)\cap I(R_2)|= |I(R_1) \Delta I(R_2)|$.
\qed
\end{proof}

\begin{definition}[Relaxed Edit distance]
    The Relaxed Edit (RE) distance between $T_1$ and $T_2$  is the TE distance under the cost function $c^* : I(R_1) \times I(R_2) \rightarrow \mathbb{R}^+$ defined as $c^*((i_1,j_1),(i_2,j_2)) =|i_1-i_2| +|j_1-j_2|$:
    $$d_{\RE}(T_1,T_2)  = d_{\TE_{c^*}}(T_1,T_2)$$
\end{definition}

\subsection{Median and Small Parsimony Problems}

We now introduce the problems investigated in the article. We consider a set of RNA trees $T_1, \ldots, T_p$ with a common leafset $L=[0,n+1].$ Before introducing the definition of the problems, we first define some constraints that can be requested for the output RNA trees of the problems.

\begin{definition}[Constraints on trees]
    An RNA tree $T$ with leafset $L$ is:
\begin{itemize}
\item Descendant-Leafset-Constrained (DLC) with respect to $T_1, \ldots, T_p$, if 
     $\DL(T) \subseteq \bigcup_{i=1}^{p}{\DL(T_i)}$

\item Internal-Leafset-Constrained (ILC) with respect to $T_1\dots T_p$, if 
     $\IL(T) \subseteq \bigcup_{i=1}^{p}{\IL(T_i)}$

\item Base-Pair-Constrained (BPC) with respect to $T_1, \ldots, T_p$, if
 $\BP(T) \subseteq \bigcup_{i=1}^{p}{\BP(T_i)}$
 
 \item Not constrained (NC) if  no constraint is applied.
\end{itemize}

    \end{definition}

Note that by definition, any tree $T_i, 1\leq i \leq p$ is DLC, ILC, and BPC with respect to $T_1, \ldots, T_p$.

\begin{problem}[\sc D\_C Median Problems]~\\
\textbf{Input}: A set of RNA trees $T_1, \ldots, T_p$ with common leafset $L$; A  distance measure $D \in \{d_{\RF}, d_{\IL}, d_{RE}\}$; A constraint on trees $C \in \{NC, DLC,ILC,BPC\}$ with respect to $T_1, \ldots, T_p$. \\
    \textbf{Output}: An RNA tree $T$ with leafset $L$, satisfying the constraint $C$, if non-empty, and minimizing
    $$\text{\tt Mcost}(T)=\sum_{i=1}^p D(T_i,T)$$
    \label{median_pb}
\end{problem}

\vspace{-.7cm}
\begin{problem}[\sc D\_C Small Parsimony Problems]~\\
\textbf{Input}:  A set of RNA trees $T_1, \ldots, T_p$ with common leafset $L$; A distance measure $D \in \{d_{RF}, d_{IL}, d_{RE}\}$; A constraint on trees $C \in \{NC, DLC,ILC,BPC\}$ with respect to $T_1, \ldots, T_p$; A phylogenetic tree $\mathbb{T}$ with leafset $\{1,\dots,p\}$ such that each leaf $i$ is assigned the tree $T_i$.\\
\textbf{Output}: A set of RNA trees $\{T_x\mid x\in I(\mathbb{T})\}$ assigned to internal nodes of $\mathbb{T}$, each satisfying the constraint $C$, if non-empty, and minimizing
$$\text{\tt SPcost}(\{T_x\mid x\in I(\mathbb{T})\}) = \sum_{(u,v)\in E(\mathbb{T})}
D(T_u,T_v)$$

\label{small_parsimony_pb}
\end{problem}
Note that in the D\_C Small Parsimony problem, we do not require the input phylogeny $\mathbb{T}$ to be binary.

\vspace{-.5cm}
\section{Computational Complexity of the D\_C Medians}
\label{algo_median_pb}

%\scriptsize{
 %% \begin{tabular}{|c|c|c|c|c|}
 %%    \hline
    
 %%         D $\backslash$ C  & NC & DLC & ILC & BPC \\
 %%    \hline
 %%        $d_{RF}$ &   P & P & P & ? \\
 %%    \hline
 %%        $d_{IL}$ &  ? & ? & P & ? \\
 %%    \hline
 %%        $d_{RE}$ &  ? & ? & ? & ? \\
 %%        \hline
 %%    \end{tabular}
%}

Given an RNA tree $T$ with an ordered leafset $L=[0,n+1]$, any descendant leafset of an internal node $x$ of $T$ is an interval of $L$, i.e., $\DL(x)=[i,j]$ with $0\leq i < j \leq n+1$. Likewise, any internal leafset of an internal node $x$ is a subset of $L$. We denote by $\DL(L)$ the set of all intervals of $L$, and by $\IL(L)$ the set of all subsets of $L$. 
%We typically refer to any element of $\DL(L)$ as a descendant leafset (DL), and any element of $\IL(L)$ as an internal leafset (IL). 
We say that an RNA tree $T$ \emph{displays} a given DL $X$ if there is a node $x$ of $T$ such that $\DL(x)=X$. Similarly, $T$ \emph{displays} a given IL $I$ if there is a node $x$ of $T$ such that $\IL(x)=I$. Finally, $T$ displays a given base pair $(i,j)$ if $(i,j)$ is an internal node of $T$.

\vspace{-.1cm}
\subsection{RF\_C Medians}

The main idea behind the computational and algorithmic results for the Median Problems under the RF distance is that, given an ordered leafset $L=[0,n+1]$, two DLs $X$ and $Y$ in $\DL(L)$ can be conflicting. Formally, $X$ and $Y$ are conflicting if they intersect, and their symmetric difference contains elements of both intervals. In this case, $X$ and $Y$ cannot be the descendant leafsets of two internal nodes of an RNA tree $T$.

The following property of conflict-free descendant leafsets is well-known for general rooted trees (see e.g.,~\cite{semple2003phylogenetics}). It still holds for RNA trees.

\begin{proposition}[Conflict-free descendant leafsets]\label{prop:rf-conflict-free}
Given an RNA tree  $T$, $\DL(T)$ is conflict-free, i.e any pair of descendant leafsets in $\DL(T)$ are not conflicting. Conversely, for any conflict-free subset $\DL'$ of  $\DL(L)$ such that $L\subseteq DL'$, one can find an RNA tree $T$ such that $DL'=\DL(T)$.
\label{conflict_free_prop}
\end{proposition}

Computing the median of a set of general rooted trees  under the RF distance has already been solved in \cite{barthelemy1986median}. The solution is also valid for RNA trees.

\begin{proposition}[Majority-rule consensus tree]
Let $\DL^+(T_1, \ldots, T_p)$ be the set of DLs
displayed by strictly more than than half of a set of RNA trees $T_1, \ldots, T_p$ with common leafset $L$. There exists an RNA tree $T^*$ such that $\DL(T^*) = \DL^+(T_1, \ldots, T_p)$ that is an optimal solution  of the RF\_NC Median Problem for  $T_1, \ldots, T_p$. 
\label{majority_rule}
\end{proposition}

The proof of Proposition~\ref{majority_rule} can be found in Appendix.
Note that Proposition~\ref{majority_rule} also solves the RF\_DLC and RF\_BPC Median Problems, as all DLs displayed by $T^*$ are displayed by at least one tree $T_i$, and thus all base pairs displayed by $T^*$ are also displayed by at least one tree $T_i$.

\subsection{IL\_C Medians}

The majority-rule idea does not appear to be applicable to the IL Median Problems. Unlike DLs, there is no guarantee that one can construct an RNA tree that contains \emph{exactly} the ILs that occur in more than half the trees. We thus develop a novel dynamic programming approach.

Note that an IL $I\in \IL(L)$ can always be expressed as a union of maximal intervals of $L$. We call \emph{gaps} of $I$ the set of intervals $[i, j]$ of $L$ such that none of $[i, j]$ belongs to $I$, but $i-1$ and $j+1$ belong to $I$ (note that these are uniquely defined). We denote by $\Gamma(I)$ the set of gaps of $I$. For integers $i, j$, we further define $\mathcal{I}{i,j} = \{ I \in \IL(L) | I \subseteq [i, j] ~\text{and} ~ I \in \bigcup{k=1}^p \IL(T_k) \}$, that is, the set of ILs contained in $[i, j]$ that occur in at least one input tree.
Two distinct ILs $I$ and $J$ are \emph{conflicting} if either $I\cap J\neq \emptyset$ or $\exists {i,j}\subseteq I$ and ${k,l}\subseteq J$ such that $i < k < j < l$ or $k < i < l < j$.
If $I$ and $J$ are not conflicting, and $\min(I)<\min(J)<\max(J)<\max(I)$, then all elements of $J$ must be included in the same gap of $I$. We write $I \preccurlyeq_{IL} J$ when this is the case. Note that $\preccurlyeq_{IL}$ defines a partial order on $\IL(L)$.

We now define the concept of \emph{structural partitions} of $[i, j]$, with $0 \leq i < j \leq n + 1$, which are the partitions of $[i, j]$ that are in bijection with RNA trees with leafset $[i, j]$.

\begin{definition}[Structural partitions]
    A structural partition of $[i, j]$ is a partition of $[i, j]$ such that:
    \begin{itemize}
        \item each set in the partition is of size at least $2$;
        \item the sets in the partition are not conflicting.
    \end{itemize}
\end{definition}

The following Lemma establishes a one-to-one correspondance between
structural partitions and RNA trees (see proof in Appendix).

\begin{lemma}  \label{lem:il_structural_partition}
Given an RNA tree $T$ with leafset $[i, j]$, the set  $\IL(T)$ is a structural partition of $[i, j]$.
Conversely, given a structural
partition $\mathcal{I}$ of $[i, j]$ such that $i$ and $j$ are in the same set of the partition, there exists a single RNA tree $T$ with leafset $[i, j]$ such that $\IL(T)=\mathcal{I}$. 
\end{lemma}

We next rewrite the cost function of a median, which will make it easier to design dynamic programming recurrences.

\begin{lemma}[Rewriting cost function]
\label{lem:il_score_function_rewriting}
An RNA tree $M$ is an IL median if and only if it minimizes:
    $$\sum_{I\in IL(M)} \left(|\{i\mid I\notin \IL(T_i)\}| - |\{i\mid I\in \IL(T_i)\}|\right)$$
\end{lemma}
%moved to appendix
% \begin{proof}
%     By definition, $M$ must minimize $\sum_{1\leq i\leq p} |\IL(T_i)\Delta\IL(M)|$.
%     It can be rewritten as such:
%     \begin{align*}
%         \sum_{i=1}^p |\IL(T_i)\Delta\IL(M)| &= \sum_{i=1}^p (|\IL(T_i)\setminus \IL(M)|+|\IL(M)\setminus \IL(T_i)|) \\
%         &= \sum_{i=1}^p (|\IL(T_i)|-|\IL(T_i)\cap\IL(M)|)~+ \\
%         &~\sum_{I\in \IL(M)} |\{ i | I \notin IL(T_i) \}| \\
%         &= \sum_{i=1}^p |\IL(T_i)|-\sum_{I\in \IL(M)} |\{i\mid i\in \IL(T_i)\}|~+ \\
%         &~\sum_{I\in\IL(M)} |\{i\mid I\notin \IL(T_i)\}|
%     \end{align*}
% Given that $\sum_{i=1}^p |\{\IL(T_i)\}|$ does not depend on the choice of $M$, finding a median is equivalent to minimizing the expression as stated.
% \qed
% \end{proof}
For conciseness, for IL $I$ we write $cost_{IL}(I)=|\{i\mid I\notin\mathcal{IL}(T_i)\}| - 
|\{i\mid I\in \IL(T_i)\}|$.
Note that the cost can be negative.
The idea of the proof is that by the properties of the symmetric difference, the median must be penalized for its ILs not in the input trees (first term in the summation), but encourages the creation of ILs that are in those trees (second term in the summation).
The main advantage of this rewriting is that it allows to break down the cost of a median as a sum of costs of individual ILs. 
That is, by Lemma~\ref{lem:il_structural_partition}, it suffices to find a structural partition of $[0, n+1]$ of minimum sum-of-costs.  
% This allows using a dynamic programming strategy, in which sub-problems
% are indexed by intervals of the  common ordered leafset. In this sense,
% it is akin to the traditional RNA folding algorithms~\citep{nussinov1980fast}. 

\vspace{-.1cm}
\subsubsection*{Algorithm for the IL\_ILC Median Problem}

Recall that in the IL\_ILC Median Problem, each IL of the median must be in the input.  To this end, for $i, j \in [0, n+1]$, we define
\begin{align}
c[i,j] = \max_{{\footnotesize\begin{tabular}{c}
$S\subseteq \mathcal{I}_{i,j}$, \\
    \text{structural}   \\
      \text{partition of }$[i,j]$
\end{tabular}}}  \sum_{I\in S} cost_{IL}(I) 
    \label{eq_dp_table_def}
\end{align}
Note that we require $S\subseteq \mathcal{I}_{i,j}$ because of the ILC constraint.  
Observe that $c[0, n+1]$ represents the cost of an IL\_ILC median (rewritten according to Lemma~\ref{lem:il_score_function_rewriting}).  
% Note that we do not require $0, n+1$ to be in the same set as in Lemma~\ref{lem:il_structural_partition}, but in the detailed proof we show that this holds since every input RNA structure has this property.

We define $c[i, j] = \infty$ if $\mathcal{I}_{i,j}$ is empty (in particular when $i \leq j$).  For the more general case, 
of course, there are too many structural partitions to enumerate.  The recursive computation of the entries of $c$ relies on the fact that exactly one set $I$ in the structural partition $S$ must contain $i$.  Since we constrain $I$ to have internal leafset in $\mathcal{I}_{i,j}$, it suffices to enumerate in polynomial time every such possible IL, and to optimize each gap $[x, y]$ independently (since each other leafset must fit within one of these gaps).  Note that if $\max(I) < j$, then we must also optimize over the remaining gap $[\max(I)+1, j]$ (which could be empty if $k = j$).
This leads to the recurrence: 
\begin{align*}
c[i,j] = \min_{\footnotesize\begin{tabular}{c}
     $I \in \mathcal{I}_{i,j}$\\
     $\text{s.t } i \in I$
\end{tabular}} \left[ cost_{IL}(I) + \sum_{[x, y] \in \Gamma(I) \cup \{[\max(I) + 1, j]\}} c[x, y] \right] 
\end{align*}
By computing $c[i, j]$ in increasing order of $j - i$, we eventually obtain $c[0, n+1]$, which is the minimum cost of a structural partition.  We can obtain a median using a standard recursive backtracking procedure.

\subsubsection*{Generalization of the algorithm to solve the IL\_NC Median Problem}

We next  now show that the dynamic programming scheme  may be generalized to the unconstrained IL\_NC median.
We define $\hat{c}[i, j]$ as the (unconstrained) structural partition of $[i, j]$ of minimum cost, namely:
\begin{align*}
\hat{c}[i,j] = \min_{{\footnotesize\begin{tabular}{c}
$S$, \text{structural}   \\
      \text{partition of }$[i,j]$
\end{tabular}}}  \sum_{I\in S} cost_{IL}(I) 
    \label{eq_dp_table_def2}
\end{align*}

In order to enforce a minimum size of $2$ for every leaf-set,
we set $\hat{c}[i,i]=+\infty$, $\forall i \in[0,n+1]$. Additionally, 
for compactness of exposition of the recursive equations, we define $\hat{c}[i,j]=0$ if $j<i$.

The main difficulty in computing $\hat{c}[i, j]$ is that with the ILC constraint, there were few possible ILs in $\mathcal{I}{i,j}$ to enumerate for an entry $c[i, j]$, but there are exponentially many possibilities after lifting the constraint.
The main idea is to consider two cases: either the IL $I$ that contains $i$ is in $\mathcal{I}{i,j}$, or not. We already know how to handle the former case.
In the case that $I$ is not in any input tree, observe that every $I$ that is not in $\mathcal{I}{i,j}$ has $cost{IL}(I) = p$, the number of input trees.
As we know that $|I|\geq 2$, we search for the smallest $k$ that is part of the same IL as $i$. For such a $k$, we can use $\hat{c}[i+1, k-1]$ to optimize between $i$ and $k$ and optimize the gaps in $[k+1, j]$. There may be elements in $[k+1, j]$ that are in the same IL as $i$ and $k$, so we cannot simply invoke $\hat{c}[k+1, j]$ as this entry ignores the existence of $i$ and $k$. Instead, we search for the optimal way to have gaps in the interval $[k+1, j]$.

This generalization, therefore, relies on a polynomial-time subroutine solving the \emph{maximum weighted independent set on interval graphs}. This subroutine will be called on a family of vertex-weighted graphs denoted $G_{k, l}$. More explicitly, $G_{k, l}$ has a vertex for every sub-interval $[u,v]$, where $k< u<v< l$.
Two sub-intervals $[u,v]$ and $[x,y]$ are connected by an edge in $G_{k, l}$ if they overlap, i.e. if $[u, v] \cap [x, y] \neq \emptyset$. The weight associated with a vertex $[u,v]$ in $G_{k, l}$
is $-\hat{c}[u,v]$, which is minus one times the minimum achievable cost over interval $[u, v]$
(we multiply by $-1$ because the independent set is a maximization problem). An independent set of $G_{k, l}$ is a set of vertices that do not share any edge. The maximum weight of an independent set of $G_{k,l}$ is denoted $\alpha(G_{k,l})$.

As mentioned above, we minimize over the case where $I$ is assumed to be in $\mathcal{I}_{i,j}$ or not, which is computed in temporary entries $\hat{c}_1[i, j], \hat{c}_2[i, j]$, respectively, as follows:
\vspace{-.4cm}
\begin{align*}
\hat{c}_1[i, j] &= \min_{\footnotesize\begin{tabular}{c}
     $I \in \mathcal{I}_{i,j}$\\
     $\text{s.t } i \in I$
\end{tabular}} \left[ cost_{IL}(I) + \sum_{[x, y] \in \Gamma(I) \cup [\max(I) + 1, j]} \hat{c}[x, y] \right]  \\
\hat{c}_2[i, j] &= \min_{i < k \leq j} (p + \hat{c}[i+1,k-1] - \alpha(G_{k+1,j})) \\
\hat{c}[i,j] &= \min (\hat{c}_1[i, j], \hat{c}_2[i, j])
\end{align*}
For the computation of $\hat{c}_2[i,j]$, $p$ represents the cost of the IL that contains $i$ and $k$, $\hat{c}[i+1,k-1]$ is the cost of the gap between $i$ and $k$, and $-\alpha(G_{k+1,j})$ the cost of the gaps to the right of $k$.

Algorihm~\ref{c2_dp_algorithm}, which can be found in Appendix, shows the pseudo-code to compute $c$ and $\hat{c}$, as well as the backtracking procedure to reconstruct an actual solution.  The unconstrained case relies on a linear-time algorithm for independent sets in interval graphs~\citep{hsiao1992efficient}.

%\bm{TODO: either in this theorem, or in another, smaller proposition 
%(which ever is quicker to write), argue that a change of score
%function in this algorithm solves RF\_ILC}
%\ml{Je pense que ça serait suffisant d'ajouter quelques phrases informelles après le théorème.  Ça ou ajouter une Prop avec un argument informel, je n'ai pas de préférence.}
\begin{theorem}\label{thm:recurrences} 
    An IL\_ILC median  can be found in time 
    in $O(p\cdot n^3)$, where $p$ is the number of input trees and $n$ is the number of leaves in the input trees.
    Moreover, an unconstrained IL median can be found in time $O(p \cdot n^5)$.
\end{theorem}

In fact, 
Theorem~\ref{thm:recurrences} is valid for the RF\_ILC median problem under a slight
change of the cost function (that is, minimizing the RF distance while requiring every IL to be in some input tree). Specifically, if we define:
\vspace{-.1cm}
\begin{align*}
cost_{RF\_ILC}(I) =  \sum_{I\in \IL(M)} (&|\{i \mid [\min(I),\max(I)] \notin \DL(T_i)\}|~-\\ &|\{i \mid [\min(I),\max(I)]\in \DL(T_i)\}|)
\end{align*}
The result of Lemma~\ref{lem:il_score_function_rewriting} holds for 
the RF distance. Therefore, defining the dynamic programming table
$c$ with this new cost function optimizes for the RF distance. 
Changing the cost function
within the recurrences has no impact on the algorithmic complexity.

\subsection{RE\_C Medians}
The approaches used for computing the median under the RF distance and the IL distance cannot be applied to the RE distance. For the RF distance, we could use the majority-rule approach because we could determine the set of input trees displaying a given DL, independently of all other DLs. Similarly, for the IL distance, we could determine the set of input trees displaying a given IL, independently of all other ILs. This allowed us to rewrite the cost function of a median tree as a sum of independent costs for each IL (Lemma \ref{lem:il_score_function_rewriting}). However, for the RE distance, different base pairs can be mapped, yielding a local cost greater than $0$. Therefore, the mapping of a given base pair in a median tree depends on the mapping of other base pairs in the tree, and thus the cost cannot be expressed as a sum of independent costs for each base pair. This limitation prevents the use of dynamic programming algorithms that build the median tree progressively. Yet, the complexity of the RE\_C Median Problems remains an open problem.

\vspace{-.5cm}
\section{Computational Complexity of D\_C Small Parsimony}
\label{algo_small_parsimony_pb}

% \scriptsize{
%  \begin{tabular}{|c|c|c|c|c|c|c|c|c|}
%     \hline
    
%          D $\backslash$ C  & $\emptyset$ & DLC & ILC & BPC & DLC,ILC & DLC,BPC  & ILC,BPC & DLC,ILC,BPC\\
%     \hline
%         $d_{RF}$ &   $\emptyset$ & DLC & ILC & BPC & DLC,ILC & DLC,BPC  & ILC,BPC & DLC,ILC,BPC\\
%     \hline
%         $d_{IL}$ &  $\emptyset$ & DLC & ILC & BPC & DLC,ILC & DLC,BPC  & ILC,BPC & DLC,ILC,BPC\\
%     \hline
%         $d_{RE}$ &   $\emptyset$ & DLC & ILC & BPC & DLC,ILC & DLC,BPC  & ILC,BPC & DLC,ILC,BPC\\
%         \hline
%     \end{tabular}
% }

\subsection{RF\_C Small Parsimony}

We give here a polynomial-time
algorithm for the unconstrained {\sc RF\_NC Small Parsimony}.  The algorithm is largely inspired by the one from~\cite{feijao2011scj}, which solves the
small parsimony for
genome rearrangements under the SCJ distance.  
Interestingly, when genomes are specified as lists of \emph{adjacencies}, they also have a notion of \emph{conflict} that applies in the same manner as DLs apply to trees.  That is, a set of adjacencies corresponds to a genome if and only if they are conflict-free.  This is analogous to the fact the a set of DLs corresponds to a tree if and only if they are conflict-free. 
We remark that the algorithm given in~\cite{feijao2011scj} does
not depend on the particular notion of conflict they use, and can therefore be adapted to our case.  Although the authors showed how to solve the Small Parsimony Problem on binary trees, we show that it can be extended to non-binary trees.  The idea is to view each possible DL as present (1) or absent (0) in each leaf, and to execute the Fitch-Hartigan algorithm on the $0-1$ representation of the DLs (see~\citep{Fitch_1971, hartigan1973minimum}).  As it turns out, we can avoid conflicts by choosing a $0$ at the root  whenever the algorithm gives us a choice between $0$ and $1$.

Recall that our input is a set of trees $T_1, \ldots, T_p$, and phylogeny $\mathbb{T}$ with $p$ leaves, where the $i$-th leaf is assigned to the RNA tree $T_i$.  
In what follows, we will denote $\mathcal{DL} = \bigcup_{i=1}^p DL(T_i)$ as the set of DLs that occur in at least one tree. 
For any such $c \in \mathcal{DL}$ and node $u \in V(\mathbb{T})$ of the input
phylogeny, we define the
bottom-up sets $B(c,u)$ in the following way:
\begin{itemize}
    \item If $u$ is a leaf assigned to tree $T$, $B(c,u)=\{1\}$ if $c \in DL(T)$, and $B(c,u)=\{0\}$ otherwise.
    \item If $u$ is not a leaf, let $children(u)$ be the set of children of $u$ in $\mathbb{T}$.  Denote by $N_0(c, u)$ (resp. $N_1(c, u)$) the number of children $v$ of $u$ satisfying $B(c, v) = \{0\}$ (resp. $B(c, v) = \{1\}$). 
    % \begin{align*}
    %     N_0(c, u) &= |\{v \in children(u) | B(c, v) = \{0\} \}| \\ 
    %     N_1(c, u) &= |\{v \in children(u) | B(c, v) = \{1\} \}|.
    % \end{align*}
    Then:
        \begin{itemize}
            \item 
            if $N_0(c, u) > N_1(c, u)$, then $B(c, u) = \{0\}$\;

            \item 
            if $N_1(c, u) > N_0(c, u)$, then $B(c, u) = \{1\}$\;
            \item 
            otherwise, $B(c, u) = \{0, 1\}$.
        \end{itemize}
\end{itemize}

Note that $B(c, u)$ is as computed in the Fitch-Hartigan approach for binary characters.  As stated in~\cite[Theorem 5.2.1]{semple2003phylogenetics}, $B(c, u)$ contains $1$ (resp $0$) if some optimal solution of the subtree rooted at $u$ puts $c$ (resp. does not put $c$) in the RNA tree inferred at $u$.  
Once the bottom-up sets have been built, final assignments
$F(c,u)$, with $c \in C$ and $u \in V(\mathbb{T})$ a node of the input phylogenetic tree, are computed
in the following way:
\begin{itemize}
    \item If $u$ is the root, and $0\in B(c,u)$ then $F(c,u)=0$ . Otherwise, $F(c,u)=1$.
    \item If $u$ is not the root, then let $w$ be its parent. If $B(c,u)=\{0,1\}$
        then $F(c,u)=F(c,w)$. Otherwise, we have $B(c,u) = \{x\}$ for $x=0$ or $1$, and we put $F(c,u)=x$.
\end{itemize}
Again, this is the same as in the Fitch-Hartigan method, except that we give a priority to $0$ if there is a choice at the root (in which case the choice propagates to the descendants as long as they have $0$ in their possible assignment).
We adapt here the Lemmas of~\cite{feijao2011scj} --- since the proofs are very similar, they are relegated to the appendix.  The main difference is that we deal with DLs instead of adjacencies, and we also deal with a possibly non-binary tree $\mathbb{T}$, but these differences are marginal.  The first step is to argue that candidate states enforced by $B$ cannot be conflicting, and the second step that the final assignments $F(c, u)$ are also free of conflicts.

\begin{lemma}\label{lem:bs-are-fine}
    Let $c, d \in C$ be two DLs in conflict and let $u$ be a node
    of $\mathbb{T}$.  If $B(c,u)=\{1\}$ then $B(d,u)=\{0\}$.
\end{lemma}
% \begin{proof} By induction on the height of the tree.
%     If $u$ is a leaf, then the property is true because an RNA tree (i.e.
%             a conflict-free set of DLs) was assigned to $u$ as input.
    
%     Suppose that $u$ is an internal node, and that the property is true for every child of $u$.  
%         Assume that $B(c,u)=\{1\}$.
%         Then $N_1(c, u) > N_0(c, u)$, that is with respect to $c$, $u$ has strictly more children with label $\{1\}$ than label $\{0\}$.  

%         Let $v$ be a child of $u$ with $B(c, v) = \{1\}$.  By induction, we know that $B(d, v) = \{0\}$.  That is, each $\{1\}$ child of $u$ with respect to $c$ has a $\{0\}$ with respect to $d$, which lets us deduce that $N_0(d, u) \geq N_1(c, u)$.
%         Likewise, let $v$ be a child of $u$ with $B(d, v) = \{1\}$.  By induction, we have $B(c, v) = \{0\}$.  Therefore, $N_0(c, u) \geq N_1(d, u)$.  Putting the gathered inequalities together, we get 
%         \[
%             N_1(d, u) \leq N_0(c, u) < N_1(c, u) \leq N_0(d, u)
%         \]
%     and, since $N_1(d, u) < N_0(d, u)$, we put $B(d, u) = \{0\}$ as desired.
%     \qed
% \end{proof}

%The same is true for the final assignment $F(c,u)$:
\begin{lemma}\label{lem:fs-are-fine}
Let $c, d \in C$ be conflicting DLs and let $u$ be a node of $\mathbb{T}$. 
If $F(c,u)=1$ then $F(d,u)=0$.
\end{lemma}
% \begin{proof}
% We proceed by induction on the depth of $u$.
% As a base case, if $u$ is the root, then $F(c,u)=1$ only in the case where $B(c,u)=\{1\}$,
%         which implies by Lemma~\ref{lem:bs-are-fine}  that $B(d,u)=\{0\}$ and therefore $F(d,u)=0$.

% Suppose that $u$ is not the root but that the property holds for its parent $w$.  Assume that  $F(c,u)=1$.  
%     By the definition of $F$, this can only occur if one of the following occurs:
%         \begin{itemize}
%             \item $B(c,u)=\{1\}$, in which case $B(d,u)=\{0\}$ by Lemma~\ref{lem:bs-are-fine}, and $F(d,u)=0$ by definition;
%             \item $B(c,u)=\{0,1\}$ and $F(c,w)=1$. In this case we know                $B(d,u)\neq\{1\}$, as otherwise Lemma~\ref{lem:bs-are-fine} would imply $B(c, u) = \{0\}$.  Therefore, $0\in B(d,u)$. By our induction hypothesis, $F(c, w) = 1$ implies 
%                 $F(d,w)=0$, which in turn implies  $F(d,u)=0$.
%         \end{itemize}

%     In all cases, $F(c,u)=1$ does imply $F(d,u)=0$. \qed
% \end{proof}

Once $B$ and $F$ are computed, we can simply assign to each $u \in V(\mathbb{T})$ the RNA tree whose set of DLs are those $c \in C$ for which $F(c, u) = 1$.  This is possible since, by Proposition~\ref{conflict_free_prop}, any set of conflict-free DLs can be turned into an RNA tree.  In the Appendix, we show the following.

\begin{theorem}\label{thm:sp-rf}
    The Small Parsimony for the RF distance can be solved in time $O(pn |V(\mathbb{T})|)$, where $p$ is the number of RNA trees and $n$ their number of leaves.
\end{theorem}

\subsection{IL\_C Small Parsimony}
\label{subsec:median-based-heuristic}

The above approach for the Small Parsimony under the RF distance cannot be applied immediately to the IL distance.  Indeed, in the RF case we could use the algorithm of~\cite{feijao2011scj} because, for any set of non-conflicting DLs inferred at internal nodes, we can construct an RNA tree with exactly that set of DLs (in fact, the algorithm works for any set of objects with this property).  In the case of ILs, it is not true that for any set of non-conflicting ILs, we can build a tree with exactly those ILs.  In particular, the set of ILs of a tree must form a partition of $[0, n+1]$, which is not guaranteed by the previous algorithm.  

Since the IL median can be computed in polynomial time, we can instead make use of the popular median-based heuristic for the Small Parsimony problem \citep{sankoff1976frequency}.
That is, we can proceed as follows: (1) infer an initial set of ancestors in $\mathbb{T}$ (for example, by requiring that each ancestor is assigned one of the RNA trees that appears at the leaves); (2) for each non-root internal node $u$ of $\mathbb{T}$, find the IL median of the RNA trees assigned at the neighbors of $u$ (this includes the parent, so there are at least three neighbors); (3) if assigning the median to $u$ improves the total cost on the branches of $\mathbb{T}$, then re-assign $u$ to the median; (4) repeat until no such improvement is possible.

\subsection{RE\_C Small Parsimony}
\label{subsec:RE-sp}

The approach inspired by~\cite{feijao2011scj} cannot be applied to the RE\_C Small Parsimony either because the Fitch-Hartigan method cannot be applied to compute a minimum presence/absence (0/1) assignment for each base pair, independently of other base pairs. Indeed, two different base pairs from two distinct input trees can be mapped in an optimal assignment of RNA trees to ancestral nodes. Therefore, base pairs cannot be considered independently of each other. The complexity of the RE\_C Small Parsimony remains open.

Since we provide no solution for the RE\_C Small Parsimony, nor for the RE\_C Median, we can only infer an assignment of RNA trees to ancestors in the phylogeny $\mathbb{T}$, with the requirement that each ancestor is assigned one of the RNA trees that appears at the leaves, and the $SPcost$ of the assignment is minimal.

\vspace{-.5cm}
\section{Experimental Results}
\label{exp_results}

\subsection{Methods} 

We implemented the algorithm underlying Theorem~\ref{thm:sp-rf} (exact solution for the RF\_NC Small
Parsimony problem) as well as the median-based heuristic described above (Section~\ref{subsec:median-based-heuristic}) for the RF\_ILC, IL\_ILC and
IL\_NC versions of the Small Parsimony problem. 
%This heuristic relies, in turn, on an implementation of Algorithm~\ref{c2_dp_algorithm}.
As initial assignment, we compute
the best combination of internal node assignments, restricted to the set
of assignments present at leaves. We also implemented the computation
of this assignment for the RE distance. In this case, it constitutes the whole heuristic (Section~\ref{subsec:RE-sp}).
All of the code is available at \url{https://github.com/bmarchand/rna\_small\_parsimony}.

%As stated in the Introduction, many heuristics for the Small Parsimony problem
%involve iteratively improving an initial assignment by computing medians. More
%precisely, the assignment at an internal node of the phylogeny is changed
%to the median of its neighbors assignments if this change improves the cost function.

%\ml{[L'heuristique de la médiane a été décrite juste en haut en section 4.2.  On peut le garder ici, mais dans ce cas on devrait supprimer 4.2 (par contre, à quelque part je garderais la justification de pourquoi l'algo de small pars ne fonctionne pas sur les IL.]}\bm{[autant pour moi, je n'avais pas vu la partie juste en haut. J'édite cette partie. (DONE)]}

\paragraph{Datasets.} These algorithms were tested on two
sets of data. 
The first one is
RFAM~(version 14.10, \cite{kalvari2021rfam}, \url{https://rfam.org/}),
the leading database for structured RNAs. It regroups RNA sequences in
families of structural homologs, for which it provides a manually-curated 
multiple alignment (seed alignment) with consensus structure annotation,
and a phylogeny. We used our algorithms to infer ancestral structures for all
internal nodes of these single-family phylogenies. 

As a pre-processing step for each
sequence in each family, a structure annotation was computed using RNAfold~\citep{lorenz2011viennarna}, under the constraint that all base-pairs
from the consensus annotation had to be kept. Note that not necessarily all base-pairs
in the consensus are valid (i.e. pair up compatible nucleotides) for all sequences
in the family. Only such valid base-pairs were kept as constraints when applying RNAfold.
Then, a gapless alignment was obtained for each family by removing from the seed alignment
any column containing a gap. When within the structural annotation of a sequence, 
a paired position is deleted this way, its partner is considered unpaired.
After this pre-processing applied, only families whose structure annotations were of length
$\leq 100$ were selected for our experiments, amounting to 2936 RNA families (out of 4170 in Rfam 14.10). Within them,
1639 families exhibit after this process the same structural
annotation for every sequence, and are therefore excluded
from our experiments.
Combined with the phylogenies provided by RFAM for each family (in which each
sequence of the seed alignment corresponds to a leaf), the
remaining 1297 families form what we will now refer to as 
the FILTERED\_RFAM dataset %\ml{[REDUCED a une connotation négative, en tout cas pour moi.  Peut-on renommer pour e.g. FILTERED? ]\bm{Oui, je suis d'accord (DONE)}}.

We also tested our algorithms on complete binary phylogenies (of height $H=5$) annotated
with random secondary structures at leaves. Structures were drawn
from the uniform distribution over the set of admissible secondary structure annotations
of length $N=100$. Formulated differently, an admissible secondary structure annotation
is a well-parenthesized string of length $N$ over the alphabet $\Sigma=\left\{(,),.\right\}$.
We add the constraint that two matching parenthesis (representing a base-pair) must
be separated by at least $\theta=3$ dots. Structures were drawn uniformly from this
set using standard Boltzmann sampling algorithms for RNA structures (\cite{ding2003statistical}, given that the uniform case corresponds to Boltzman sampling with
$\beta=0$). This constitutes what we now refer to as the RANDOM
dataset.

Note that, since the structures annotating the leaves of the FILTERED\_RFAM
dataset belong to members of the same homology family, they tend to be
more similar to each other than the structures of the RANDOM dataset. 
They therefore provide complementary settings when it comes to comparing 
our various combinations of constraint and metric.

\paragraph{Experimental protocol.} For both the FILTERED\_RFAM and RANDOM datasets,
we infer ancestral structures under the combination of metrics and
constraints RF\_NC, IL\_NC, IL\_ILC and RF\_ILC using
our polynomial exact algorithm (RF\_NC) and heuristics 
(IL\_NC, IL\_ILC and RF\_ILC, see ``median-based heuristic'' above).
Our purpose is to assess the capacity of each method to yield \emph{resolved}
ancestral structures. As a proxy for resolution, we look at the 
(average and/or maximum)
number of base-pairs of ancestral structures as we climb up the phylogenies.
The rationale is that a reasonable feature to expect from
ancestral reconstruction methods could be the conservation of the average number
of base-pairs regardless of the height in the phylogeny.
Our experimental results are displayed on Figures~\ref{fig:rfam_num_bps_height} (Appendix)
and \ref{fig:root_bps_scatter_plots}~(A) (for the FILTERED\_RFAM dataset) and
Figure~\ref{fig:root_bps_scatter_plots}~(B) (for the RANDOM dataset).

\vspace{-.4cm}
\subsection{Results and discussion}

\paragraph{Execution times.} Table~\ref{tab:exec_times} 
reports average execution times per family on the FILTERED\_RFAM dataset,
for each of our methods. 
%These values
They are consistent with the hierarchy of theoretical time complexity.  The IL\_NC approach is unsurprisingly slower and the RF\_NC faster since it does not require exploring medians until convergence.  The IL\_ILC and RF\_ILC do have to apply this step, but are reasonably fast owing to their faster dynamic programming algorithm.
The RE implementation is the slowest.

\begin{table}
\centering
\begin{tabular}{|c|c|}
\hline
   metric+constraint  & avg. time per family (seconds) \\
\hline
   RF\_NC  & 0.26 \\
\hline
IL\_NC & 10.5 \\
\hline
IL\_ILC & 1.39 \\
\hline
RF\_ILC & 1.51 \\
\hline
RE & 45.9 \\
\hline
\end{tabular}
    \caption{Average execution times per family
    of each combination of metric
    and constraint, on the FILTERED\_RFAM dataset.}
    \label{tab:exec_times}
    \vspace{-1cm}
\end{table}

\begin{figure*}
    \centering
    \includegraphics[width=.9\textwidth]{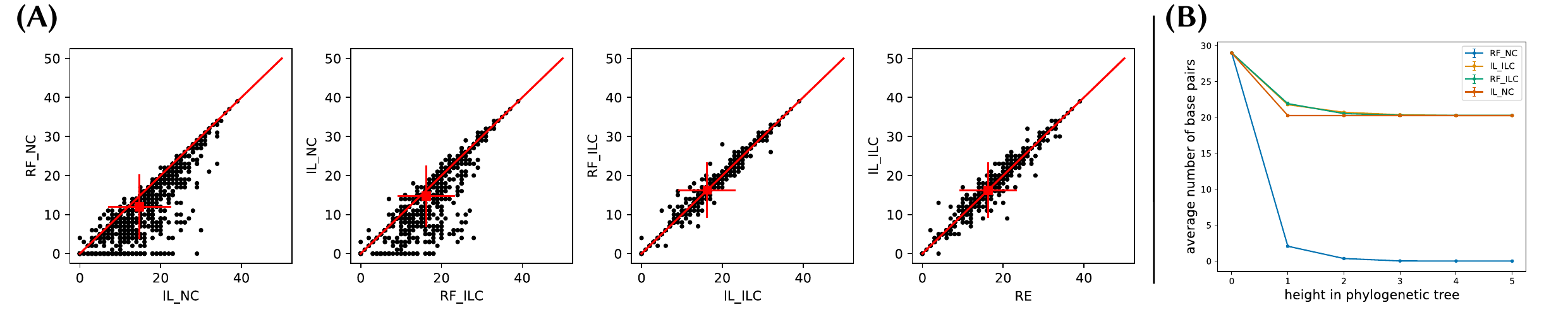}
    \caption{(A) Scatter plots of the number of base-pairs predicted
    at the root of the phylogeny by RF\_NC, 
    RF\_ILC, IL\_ILC, IL\_NC and RE for all families
    in the FILTERED\_RFAM dataset. Each point is an RFAM family,
    with its $x,y$-coordinates equal to the number of predicted
    base-pairs at the root of the phylogeny by the respective
    methods. For each plot, the red square point indicates the 
    average, its error bars being the standard deviation. 
    As in Figure~\ref{fig:rfam_num_bps_height} (Appendix),
    RF\_NC seems to provide the least resolved results.
    IL\_NC performs slightly better, but  RF\_ILC and IL\_ILC yield the most base pairs.
    Both of them seem to fare comparably with each other,
    indicating that the ILC constraint (only internal leafsets
    from the input structures) is the deciding factor. (B)
    Average maximum number of base-pairs in reconstructed ancestral
    structures, as a function of the height of the corresponding node
    in the phylogenetic tree, over the RANDOM dataset. 
    We observe as a general trend that the number of base-pairs
    in predicted ancestral structures decreases as we move
    up the trees. However, where RF\_NC very quickly
    predicts empty structures at ancestral nodes,
    the other metric/constraint combinations (RF\_ILC, IL\_ILC
    and IL\_NC) do predict non-empty structures. 
    Remarkably, IL\_NC does so without any constraint.
    }
    \label{fig:root_bps_scatter_plots}
\end{figure*}

\begin{figure*}
\centering
\includegraphics[width=.85\textwidth]{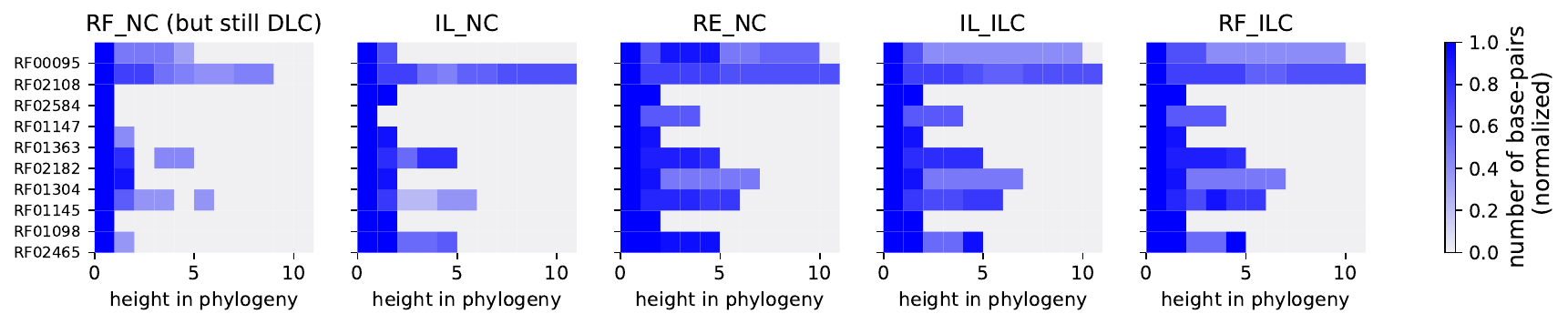}
\caption{Maximum number of base-pairs as a function
of the height of nodes in the phylogeny, for a selected set of 10 maximally-divergent RFAM families. The number of base-pairs
are normalized, for each family, by the maximum number of base-pairs
over the structures annotating the leaves.
The selected ``maximally-divergent'' families are the ones maximizing the sum of distances over pairs of leaves, as measured
by the Internal-Leafset distance (Definition~\ref{def:il_distance}).
Solving Small Parsimony under the metric/constraint combination
RF\_NC tends to yield ancestral structures with few
base-pairs, as we move up the phylogenies. While also unconstrained,
IL\_NC tends to predict more base-pairs than RF\_NC in ancestral structures. Being constrained
to use only internal-leafsets from the input structures, 
IL\_ILC and RF\_ILC predict the most resolved ancestral structures,
as per the criteria of the number of base-pairs. The
score function difference (IL vs. RF) does not seem to have
more than marginal impact. Note that RF\_NC is DLC (only descendant leaf-sets from the input structure) so imposing this constraint would not help get more resolution.
}
\label{fig:rfam_num_bps_height}
\end{figure*}

\paragraph{Resolution of ancestors.}Figures~ \ref{fig:root_bps_scatter_plots}~(A), \ref{fig:root_bps_scatter_plots}~(B), and ~\ref{fig:rfam_num_bps_height} (Appendix)
aim at representing the effect of the choice
of metric and constraint on the \emph{resolution} of the Small Parsimony problem.
By resolution, we mean the tendency to infer ancestors with detailed characteristics,
in spite of divergences between leaf annotations. 
Within this work, 
we use the number of predicted base-pairs as a proxy for the resolution of ancestral structures.

\paragraph{Low resolution for RF\_NC.} The Robinson-Foulds metric is a popular
choice for comparing phylogenetic trees~\citep{robinson1981comparison}. When it comes to RNA structures however, our results on Figures~\ref{fig:rfam_num_bps_height} (Appendix),
\ref{fig:root_bps_scatter_plots}~(A) and \ref{fig:root_bps_scatter_plots}~(B) 
show that solving Small Parsimony under this distance leads to
reconstructed ancestral structures with few base-pairs. 
The results are particularly striking for the RANDOM dataset (Figure~\ref{fig:root_bps_scatter_plots}~(B)). This makes sense in the light of the fact that in the RANDOM dataset, structures at leaves tend to be more 
distant to each other, compared to FILTERED\_RFAM dataset.
This low resulution of RF\_NC 
might be partially due to the systematic preference of 0 (absence of
a descendant leaf-set) over 1 (presence) at the root of the phylogenies or, perhaps more likely, to the absence of constraints.  In particular,
as formulated in Proposition~\ref{conflict_free_prop}, \emph{any} set of 
non-conflicting descendants leaf-sets yields a valid structure. If this set
is small, the corresponding structure has few base-pairs.

\paragraph{Intermediate resolution for IL\_NC.}
Figure~\ref{fig:root_bps_scatter_plots} and Figure~\ref{fig:rfam_num_bps_height} (Appendix) show that for the FILTERED\_RFAM dataset, intermediate levels of resolution are achieved with IL\_NC (higher than RF\_NC, lower than
IL\_ILC and RF\_ILC). This may be due to the fact that the IL distance
favors the presence of \emph{loops} from the input structures, and a loop is present only if \emph{all the base-pairs defining it} are present.

\paragraph{Highest resolution with the ILC constraint.} The highest
resolution is obtained by enforcing the ILC constraint (only
internal leaf-sets from the input structures). 
A possible explanation for this fact is that, in the case of high divergence between some of the leaf solutions, the best median
solution may be an empty structure. The ILC constraint forbids this
possibility, and enforces higher resolution at ancestral nodes.
This occurs regardless
of the choice of metric (RF or IL), signaling the strength of the 
ILC constraint. As for the RE method, given its restriction (Section~\ref{subsec:RE-sp})
to leaf annotations for internal nodes, it unsurprisingly yields high resolution.

%\begin{figure}
%\centering
%\includegraphics[width=.45\textwidth]{figures/average_num_bps_height_random_structures.pdf}
%    \caption{
%    }
%    \label{fig:random_structures_results}
%\end{figure}

\begin{figure}
    \centering
    \includegraphics[width=\linewidth]{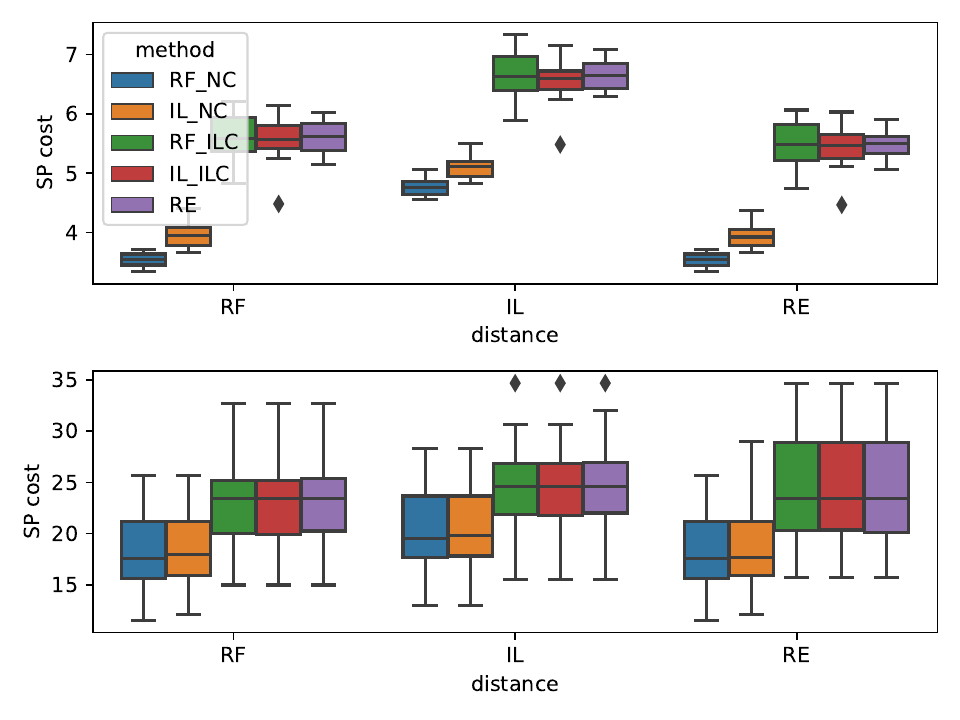}
    \caption{Distribution of the Small Parsimony (SP) costs 
    obtained over the FILTERED\_RFAM (bottom) and the RANDOM dataset with random structures of length $30$ (top), by each method with respect to all three distances.
    The Small Parsimony cost is the sum of distances (either RF, IL, or RE) over each edge of a phylogeny. 
    To allow averaging over several
    RFAM families, costs are divided by the number of edges in the
    phylogeny.}
    \label{fig:box_plot}
    \vspace{-.5cm}
\end{figure}

\paragraph{Comparison of SP costs.}
Figure~\ref{fig:box_plot} shows the distribution of SP costs (the sum of distances on the branches) for our datasets.  On the FILTERED\_RFAM dataset, we restricted this analysis to the 20 families that exhibit the most divergence at the leaves.  The unconstrained predictions typically achieve a lower SP cost than their constrained counterparts, which is not surprising since they are allowed a larger solution space.  There is little difference in terms of cost between the constrained versions.  There appears to be an inevitable trade-off: the constrained versions offer better resolution, but at the price of a higher SP cost.  On the RANDOM dataset, the differences in cost are even more pronounced.  Notably, structures inferred by the unconstrained RF distances achieve lower SP cost, again illustrating its relationship with the level of resolution.

%\appendix

%\begin{figure}
%\centering
%\includegraphics[width=.45\textwidth]{slides/tikz_figures/example_alignment.pdf}
%    \caption{Example of an alignment of structured RNAs, corresponding ``RNA block trees'' (bottom)
%    and ``edit distance trees'' (top). By ``edit distance tree'', we mean a tree built
%    from structure-annotated RNA sequence in (what seems to be ?) the ``traditional way'' of
%    doing so when using the tree edit distance on structured RNAs (see \cite{chen2014improved} for instance).}
%    \label{big_example}
%\end{figure}

\vspace{-1cm}
\section{Conclusion}
Investigating the evolution of RNA structures provides information about their functional adaptations and their role in the evolutionary history of organisms. In particular, it is expected to shed light on the RNA World hypothesis.

In this work, we have established novel foundations for the reconstruction of ancestral RNA structures by borrowing several ideas from phylogenetics. Combining research in structural RNA and phylogenetics is promising since, compared to prior work, our algorithmic framework allows comparing any set of aligned RNA secondary structures, whether the alignment displays crossing base pairs, the structures are similar or distant, homologs or not. Moreover, instead of focusing on inferring constrained ancestral RNA sequences, our approach also involves predicting the evolutionary history of the structures. Furthermore, we introduced the novel IL distance, which compares structural motifs instead of solely base pairs. To our knowledge, this distance had not been considered previously, and further exploration is required to understand its potential. Moreover, our implementation was able to assess the resolution on RFAM and randomized data.

Our work paves the way for several future research directions. In terms of algorithms, the complexity of the Small Parsimony problems under IL\_NC, IL\_ILC, and RF\_ILC is still open. The RE Median and Small Parsimony problems are also open, even for designing heuristics. We mention in passing that the IL median problem is also open for \emph{unordered} trees and may be of independent interest. In practice, it remains to improve the performance of our algorithms, especially the IL algorithms that have a cubic or quintic dependency on $n$. Moreover, we have studied the restricted version of the problems where RNA structures are not subject to nucleotide indel events, thus considering RNA sequences of the same length. Future algorithmic extensions of the problems will include considering a gapped alignment as input. This will allow our framework to handle multiple RNA families, for instance RNA families exhibiting structural differences but yet grouped into clans. Other problems of interest are to seek the optimal median or the optimal small parsimony assignment, as well as the alignment when the alignment is not provided as input.

\bibliography{biblio}
\bibliographystyle{natbib}
%\bibliographystyle{achemnat}
%\bibliographystyle{plainnat}
%\bibliographystyle{abbrv}
%\bibliographystyle{bioinformatics}
%
%\bibliographystyle{plain}
%
%\bibliography{document}

% \begin{thebibliography}{}

% \bibitem[Bofelli {\it et~al}., 2000]{Boffelli03}
% Bofelli,F., Name2, Name3 (2003) Article title, {\it Journal Name}, {\bf 199}, 133-154.

% \bibitem[Bag {\it et~al}., 2001]{Bag01}
% Bag,M., Name2, Name3 (2001) Article title, {\it Journal Name}, {\bf 99}, 33-54.

% \bibitem[Yoo \textit{et~al}., 2003]{Yoo03}
% Yoo,M.S. \textit{et~al}. (2003) Oxidative stress regulated genes
% in nigral dopaminergic neurnol cell: correlation with the known
% pathology in Parkinson's disease. \textit{Brain Res. Mol. Brain
% Res.}, \textbf{110}(Suppl. 1), 76--84.

% \bibitem[Lehmann, 1986]{Leh86}
% Lehmann,E.L. (1986) Chapter title. \textit{Book Title}. Vol.~1, 2nd edn. Springer-Verlag, New York.

% \bibitem[Crenshaw and Jones, 2003]{Cre03}
% Crenshaw, B.,III, and Jones, W.B.,Jr (2003) The future of clinical
% cancer management: one tumor, one chip. \textit{Bioinformatics},
% doi:10.1093/bioinformatics/btn000.

% \bibitem[Auhtor \textit{et~al}. (2000)]{Aut00}
% Auhtor,A.B. \textit{et~al}. (2000) Chapter title. In Smith, A.C.
% (ed.), \textit{Book Title}, 2nd edn. Publisher, Location, Vol. 1, pp.
% ???--???.

% \bibitem[Bardet, 1920]{Bar20}
% Bardet, G. (1920) Sur un syndrome d'obesite infantile avec
% polydactylie et retinite pigmentaire (contribution a l'etude des
% formes cliniques de l'obesite hypophysaire). PhD Thesis, name of
% institution, Paris, France.

% \end{thebibliography}

\newpage
\section*{Appendix}

The detailed proofs of our results can be found below.

\subsection*{Proof of Lemma~\ref{lem:il_score_function_rewriting}}

    By definition, $M$ must minimize $\sum_{1\leq i\leq p} |\IL(T_i)\Delta\IL(M)|$.
    It can be rewritten as such:
    \begin{align*}
        \sum_{i=1}^p |\IL(T_i)\Delta\IL(M)| &= \sum_{i=1}^p (|\IL(T_i)\setminus \IL(M)|+|\IL(M)\setminus \IL(T_i)|) \\
        &= \sum_{i=1}^p (|\IL(T_i)|-|\IL(T_i)\cap\IL(M)|)~+ \\
        &~\sum_{I\in \IL(M)} |\{ i | I \notin IL(T_i) \}| \\
        &= \sum_{i=1}^p |\IL(T_i)|-\sum_{I\in \IL(M)} |\{i\mid i\in \IL(T_i)\}|~+ \\
        &~\sum_{I\in\IL(M)} |\{i\mid I\notin \IL(T_i)\}|
    \end{align*}
Given that $\sum_{i=1}^p |\{\IL(T_i)\}|$ does not depend on the choice of $M$, finding a median is equivalent to minimizing the expression as stated.
\qed

\subsection*{Proof of Theorem~\ref{thm:recurrences}}

For convenience, we recall that $c[i, j]$ is the minimum cost of a structural partition of $[i, j]$ under the ILC constraint, and $\hat{c}[i, j]$ the same cost but without the constraint. 
In the main text, we provided the recurrences 
\begin{align*}
c[i,j] = \min_{\footnotesize\begin{tabular}{c}
     $I \in \mathcal{I}_{i,j}$\\
     $\text{s.t } i \in I$
\end{tabular}} \left[ cost_{IL}(I) + \sum_{[x, y] \in \Gamma(I) \cup \{[\max(I) + 1, j]\}} c[x, y] \right] 
\end{align*}
and
\begin{align*}
\hat{c}_1[i, j] &= \min_{\footnotesize\begin{tabular}{c}
     $I \in \mathcal{I}_{i,j}$\\
     $\text{s.t } i \in I$
\end{tabular}} \left[ cost_{IL}(I) + \sum_{[x, y] \in \Gamma(I) \cup [\max(I) + 1, j]} \hat{c}[x, y] \right]  \\
\hat{c}_2[i, j] &= \min_{i < k \leq j} (p + \hat{c}[i+1,k-1] - \alpha(G_{k+1,j})) \\
\hat{c}[i,j] &= \min (\hat{c}_1[i, j], \hat{c}_2[i, j])
\end{align*}

We prove here the correctness of the  recursions, and conclude with a complexity analysis.  Let $i, j$ be integers.  We show that our recurrences for $c[i, j]$ and $\hat{c}[i, j]$ correctly compute the minimum cost of a structural partition (with the ILC constraint in the former case).  

As a base case for the $c$ entries,  $\mathcal{I}_{i,j} = \emptyset$, no structural partition under the ILC constraint is possible and putting 
$c[i,j] = \infty$ is correct.  For the $\hat{c}$ entries, if $i = j$ no structural partition is possible because so set of a partition of $\{i\}$ can have two elements, and so $\hat{c}[i, j] = \infty$ is correct. 
For $i < j$, $\hat{c}[i, j] = 0$ can be seen as correct since the empty partition has cost $0$.  

Now assume inductively that $c[i', j']$ and $\hat{c}[i', j']$ are correct for any $[i', j'] \subset [i, j]$.  We argue that $c$ and $\hat{c}$ are at most the cost of an optimal structural partition, then at least its cost.

    \noindent
    ($\leq$) Let $S$ be a structural partition of $[i,j]$ of minimum cost (we handle both cases of $S$ satisfying ILC or not in the same proof). 
        Since $S$ is a partition, 
        there must be some $I \in S$ containing $i$. 
        Note that since $S$ is conflict-free, any IL other than $I$ in $S$ must be in a gap of $I$, or must be contained in $[\max(I) + 1, j]$.  Moreover, the sets of $S$ contained in such a $[x, y] \in \Gamma(I) \cup \{[\max(I) + 1, j]\}$ must form a partition of $[x, y]$.   
        
        If $S$ is required to satisfy the ILC constraint, then the cost of $S$ is at least $cost_{IL}(I)$, plus the sum of $c[x, y]$ for every such $[x, y]$ (by induction).  Since our recurrence for $c[i, j]$ minimizes over every $I \in \mathcal{I}_{i,j}$, it will, in particular, consider $I$ at some point, and thus $c[i, j]$ is no more than the cost of $S$.

        So suppose that $S$ is not required to satisfy the ILC constraint.  If $I \in \mathcal{I}_{i,j}$, by the same arguments as above, $\hat{c}_1[i, j]$ as computed above is no more than the cost of $S$, and thus the same holds for $\hat{c}[i, j] \leq \hat{c}_1[i,j]$.
        On the contrary, if $I$ is not an input leafset, then $cost_{IL}(I) = p$.  Let $k$ be the minimum integer in $I \setminus \{i\}$ (which exists since sets of $S$ have size at least two).  Then $[i+1, k - 1]$ is a gap of $I$ and the subsets of $S$ in $[i+1, k - 1]$ form a structural partition of cost at most $\hat{c}[i+1, k - 1]$, by induction.  Moreover, let $[x_1, y_1], \ldots, [x_k, y_k]$ be the other gaps of $I$, including $[\max(I) + 1, j]$ if non-empty.  Then $S$ restricted to any of these intervals forms a structural partition, and the sum of costs of these is at most $\sum_{i=1}^k \hat{c}[x_i, y_i]$.  These intervals form an independent set of $G_{k+1, j}$ with cost minus $1$ times this sum, and thus $-\alpha(G_{k+1, j})$ is no more than this sum.     
        Since the expression of $\hat{c}_2[i, j]$ minimizes over every possible $k$, $\hat{c}_2[i, j]$ is no more than the cost of $S$, and thus $\hat{c}[i, j]$ as well.
        
    \noindent
    ($\geq$) For the other direction, first consider the entry $c[i, j]$ when the ILC constraint is present.  Let $I$ be any subset of $\mathcal{I}_{i,j}$ containing $i$.  Suppose that the expression in large brackets for $c[i, j]$ is $\infty$.  Then there is some gap $[x, y] \in \Gamma(I) \cup \{[\max(I) + 1, j]\}$ such that $c[x, y] = \infty$.  Since this value is correct by induction, this means that there is no structural partition of $[x, y]$, and thus none for $[i, j]$ either when $I$ is in the partition (because it is impossible to partition one of its gaps).  
    This means that if \emph{every} $I \in \mathcal{I}_{i,j}$ yields $\infty$, there cannot be a structural partition of $[i, j]$ with the ILC condition, and we will correctly put $c[i, j] = \infty$.
    
    Assume otherwise, and suppose that $I$ minimizes the expression of $c[i, j]$. 
    For each gap $[x, y]$ of $I$, including $[\max(I) + 1, j]$ if non-empty, let $S_{x,y}$ be an optimal structural partition of $[x, y]$.  By induction, its cost is $c[x, y]$.  Taken together, $I$ and the union of the $S_{x,y}$ are easily seen to be a structural partition of cost $cost_{IL}(I)$, plus the sum of $c[x, y]$ for each gap $[x, y]$.  It thus forms a valid solution for the sub-problem associated, and may not be better than optimal, therefore our recurrence for $c[i, j]$ is at most the cost of an optimal solution for $[i, j]$.
    
    Next, suppose that the ILC constraint is not present.  Note that $\hat{c}[i, j] < \infty$.  This is because for $i < j$, the entry for $\hat{c}_2[i, j]$ can choose $k = i+1$ and choose an empty independent set, which corresponds to a solution of cost $p$.  Since this is a possibility for $\hat{c}[i, j]$, the latter is $p$ or better.
    If $\hat{c}[i, j] = \hat{c}_1[i, j]$, then one can obtain a valid structural partition of $[i, j]$ by taking the $I$ that minimizes the expression, and the optimal solution for each gap.  As in the previous case, this gives a solution of cost at most $\hat{c}_1[i, j]$.  Since it cannot be better than optimal, we get that $\hat{c}[i, j] = \hat{c}_1[i, j]$ is at most the cost of an optimal structural partition.

    So suppose that $\hat{c}[i, j] = \hat{c}_2[i, j]$.  Let $k$ be the index that minimizes its expression, and let $R$ be a maximum independent set of $G_{k+1, j}$.  Let $I = [i, j] \setminus (\{[i+1, k-1]\} \cup \bigcup_{[x, y] \in R} [x, y])$.  Note that $I$ has size at least two since it contains $i$ and $k$.  
    The partition obtained from the union of $I$ (of cost $p$), of an optimal structural partition of $[i+1, k - 1]$ (of cost $\hat{c}[i+1,k-1]$), and of an optimal structural partition of each $[x, y] \in R$ (of total cost $-\alpha(G_{k+1,j})$), yields a structural partition of $[i, j]$ of cost $\hat{c}_2[i, j]$.  Since it cannot be better than optimal, we get that $\hat{c}[i, j] = \hat{c}_2[i, j]$ is at most the cost of an optimal structural partition.

    Having completed both sides of the inequality, we see that $c[i, j]$ (resp. $\hat{c}[i, j]$) correctly computes the minimum cost of a constrained (resp. unconstrained) structural partition of $[i, j]$.
    
    To finish the argument, we need to argue that $c[0, n+1]$ and $\hat{c}[0, n+1]$ correspond to the cost of an optimal median tree.  If $0, n+1$ are in the same set of the structural partitions correspond to those $c$ and $\hat{c}$ entries, then this holds by Lemma~\ref{lem:il_structural_partition}.  It is not hard to see that some optimal structural partition must put $0, n+1$ in the same set, since the IL $\{0, n+1\}$ is in every input tree.  We omit the details.

     As for the complexity, consider the time required to compute every entry $c[i, j]$.  
     There are $O(n^2)$ entries to fill. Each of them requires iterating over every $I \in \mathcal{I}_{i,j}$ that contains $i$.  Note that each input tree has exactly one IL containing $i$, so there are $p$ entries to enumerate.  For each $I$ in this enumeration, we can compute $cost_{IL}(I)$ and the sum in time $O(n)$ (assuming constant time access to the $c[x, y]$ entries). 
     The overall complexity is therefore $O\left(n^2\cdot pn\right) = O(pn^3)$.

     Next, consider the time to calculate the $O(n^2)$ entries $\hat{c}[i, j]$.  As before, $\hat{c}_1[i, j]$ takes time $O(pn)$.  Then, $\hat{c}_2$ iterates over $O(n)$ values of $k$.  Each of them makes a call to a solver for the independent set problem over $O(n^2)$ intervals.  It is known that such an independent set can be found in time $O(n^2)$ if the input intervals are given in right-end sorted order, which is easy to construct here~\citep{hsiao1992efficient}.  Thus, the time spent for one $\hat{c}_2$ entry is $O(n^3)$.  One $\hat{c}$ entry takes time $O(pn + n^3)$, for a total time of $O(pn^3 + n^5)$.

\subsection*{Proof of Proposition~\ref{majority_rule}}

\begin{proof}
We claim that $\DL^+(T_1, \ldots, T_p)$ must be conflict-free. Indeed, given two intervals $X$ and $Y$ in $\DL^+(T_1, \ldots, T_p)$, since they are both present in strictly more than half of the input
trees, then there must be a tree $T_i$ in which they are both displayed. By Proposition~\ref{conflict_free_prop}, $X$ and $Y$ cannot be conflicting. Since $L\in \DL^+(T_1, \ldots, T_p)$, by Proposition~\ref{conflict_free_prop}, there exists an RNA tree $T^*$ such that $\DL^+(T_1, \ldots, T_p)=\DL(T^*)$. 

Now, consider $T$ an RNA tree displaying a DL $L'$ which is not displayed by more than half of the input trees, i.e. $L'\notin \DL^+(T_1, \ldots, T_p)$. Say $L'=\DL(x)$ for some internal node $x$ of  $T$. Consider now $T'$
obtained from $T$ by merging the node $x$ with its parent, effectively removing
$L'$ from the set of descendant leafsets. We then have:
$\text{Mcost}(T') = \text{Mcost}(T) + |\{i\mid L'\in \DL(T_i)\}| - |\{i\mid L'\notin \DL(T_i)\}|$
Since $L'$ is not displayed by more than half of the input trees, then $\text{Mcost}(T') \leq \text{Mcost}(T)$.  Note that the equality holds if $L'$ is displayed by exactly half of the trees. So, by merging all nodes of $T$ whose descendant leafsets are not in $\DL^+(T_1, \ldots, T_p)$ with their parents, we can obtain an RNA tree $T"$ that displays only DL that are in $\DL^+(T_1, \ldots, T_p)$ and such that $\text{Mcost}(T") \leq \text{Mcost}(T)$. In addition, if the tree $T"$ displays only a strict subset of $\DL^+(T_1, \ldots, T_p)$, i.e not all DLs in $\DL^+(T_1, \ldots, T_p)$, then 
$\text{Mcost}(T^*) < \text{Mcost}(T")$. Therefore, $T^*$ is an optimal solution of the RF\_NC Median Problem.
\qed
\end{proof}

\subsection*{Proof of Lemma~\ref{lem:bs-are-fine}}

We prove the lemma by induction on the height of the tree.
    If $u$ is a leaf, then the property is true because an RNA tree (i.e.
            a conflict-free set of DLs) was assigned to $u$ as input.
    
    Suppose that $u$ is an internal node, and that the property is true for every child of $u$.  
        Assume that $B(c,u)=\{1\}$.
        Then $N_1(c, u) > N_0(c, u)$, that is with respect to $c$, $u$ has strictly more children with label $\{1\}$ than label $\{0\}$.  

        Let $v$ be a child of $u$ with $B(c, v) = \{1\}$.  By induction, we know that $B(d, v) = \{0\}$.  That is, each $\{1\}$ child of $u$ with respect to $c$ has a $\{0\}$ with respect to $d$, which lets us deduce that $N_0(d, u) \geq N_1(c, u)$.
        Likewise, let $v$ be a child of $u$ with $B(d, v) = \{1\}$.  By induction, we have $B(c, v) = \{0\}$.  Therefore, $N_0(c, u) \geq N_1(d, u)$.  Putting the gathered inequalities together, we get 
        \[
            N_1(d, u) \leq N_0(c, u) < N_1(c, u) \leq N_0(d, u)
        \]
    and, since $N_1(d, u) < N_0(d, u)$, we put $B(d, u) = \{0\}$ as desired.

\subsection*{Proof of Lemma~\ref{lem:fs-are-fine}}

We proceed by induction on the depth of $u$.
As a base case, if $u$ is the root, then $F(c,u)=1$ only in the case where $B(c,u)=\{1\}$,
        which implies by Lemma~\ref{lem:bs-are-fine}  that $B(d,u)=\{0\}$ and therefore $F(d,u)=0$.

Suppose that $u$ is not the root but that the property holds for its parent $w$.  Assume that  $F(c,u)=1$.  
    By the definition of $F$, this can only occur if one of the following occurs:
        \begin{itemize}
            \item $B(c,u)=\{1\}$, in which case $B(d,u)=\{0\}$ by Lemma~\ref{lem:bs-are-fine}, and $F(d,u)=0$ by definition;
            \item $B(c,u)=\{0,1\}$ and $F(c,w)=1$. In this case we know                $B(d,u)\neq\{1\}$, as otherwise Lemma~\ref{lem:bs-are-fine} would imply $B(c, u) = \{0\}$.  Therefore, $0\in B(d,u)$. By our induction hypothesis, $F(c, w) = 1$ implies 
                $F(d,w)=0$, which in turn implies  $F(d,u)=0$.
        \end{itemize}

    In all cases, $F(c,u)=1$ does imply $F(d,u)=0$. 

\subsection*{Proof of Lemma~\ref{lem:il_structural_partition}}

\begin{proof}
($\Rightarrow$) Since every leaf of $T$ has a unique parent, each leaf belongs to a unique IL and this $IL(T)$ forms a partition of $[i, j]$.  Moreover, since each internal node $u$ of an RNA tree represents a base pair $(x, y)$, each set in $IL(T)$ has at least two elements. 
    
    As for the conflict-free property, consider two internal nodes
    $u,v$.  If $u$ and $v$ are incomparable (none is an ancestor of the other), then $DL(u) = [k_u, l_u], DL(v) = [k_u, l_u]$ satisfies $[k_u, l_u] \cap [k_v,l_v] = \emptyset$.  Since $IL(u) \subseteq DL(u), IL(v) \subseteq DL(v)$, no conflict is possible.  
    So suppose w.l.o.g that $u$ is an ancestor of $v$.  Since $T$ is an ordered tree, all leaves that descend from $v$ (including $IL(v)$), are contained in the same gap of $IL(u)$, from which it is easy to see that no conflict is possible.
    
    % , and the list of nodes of $T$ according to a depth-first-search exploration
    % of the tree following the ordering of the children at each node (with w.l.o.g $u$ being visited before $v$).  Note that the leaves are enumerated in order $i, i+1, \ldots, j$.
    % Suppose that $u$ is not an ancestor of $v$.  Note that $\min (IL(v)) > \max (IL(u))$ since $v$ is enumerated after $u$ and all of its descendants.  This implies that $i < k$ for every $i \in IL(u)$ and $k \in IL(v)$, and hence no conflict is possible. 
    % So suppose that $u$ is an ancestor of $v$.  Then there is a
    % child $w$ of $u$ which is an ancestor of every element of $IL(v)$.  Every element $k \in IL(u)$ is therefore smaller or greater than every element that descends from $w$, which means that either $k < \min IL(v)$ or $k > \max IL(v)$ for every such $k$.  This implies that $IL(u)$ cannot intersect with $IL(v)$ and that there cannot be $i', j' \in IL(u)$, $k', l' \in IL(v)$ with $i' < k' < j' < l'$ or $k' < i' < l' < j'$.
    \noindent
($\Leftarrow$) Conversely, let $\mathcal{I}$ be a structural
    partition of $[i, j]$, with $i, j$ in the same set of $\mathcal{I}$.  
    An RNA
    tree can be obtained by creating an internal node for each element of $\mathcal{I}$, 
    creating a leaf for each element $x$ in $[i,j]$, and connecting $x$ the unique internal node whose
    IL contains $x$. A node associated to IL $I \in \mathcal{I}$ is made the parent of a node associated to IL $J \in \mathcal{I}$ if $I\preccurlyeq J$
    and there is no $I'$ such that $I\preccurlyeq I' \preccurlyeq J$.
    It is not hard to see that this creates a tree rooted at the IL that contains $i, j$, and that the conflict-free property allows ordering the children of nodes as required by RNA trees.
\qed
\end{proof}

\subsection*{Proof of Theorem~\ref{thm:sp-rf}}

Let us first mention that the $F$ assignments produce an optimal solution.  
Indeed, because the set of assigned DLs are those obtained from the Fitch-Hartigan approach, we know that $F$ yields a solution that minimizes the $0-1$ or $1-0$ changes on the branches, which is equivalent to minimizing the symmetric difference.  We refer to~\citep{semple2003phylogenetics} and~\citep{feijao2011scj} for more details on the optimality of the approach.

In terms of complexity, for each $c \in \mathcal{DL}$, computing one $B(c, u)$ entry requires iterating over the children of $u$.  Thus, the time needed to compute all $B(c, u)$ entries is proportional to the sum of number of children of nodes $u$ in $\mathbb{T}$, which is $O(|V(\mathbb{T})|)$.  The time to compute all the $F$ entries is no more than for $B$.  Thus, the total time for the $B$ and $F$ phase is $O(|C| |V(\mathbb{T})|)$, which is $O(pn |V(\mathbb{T})|)$ (with $p$ the number of RNA trees and $n$ the number of leaves).  For each $u \in V(\mathbb{T})$, we can reconstruct the RNA tree to assign to it in time $O(pn)$ by listing the DLs $c$ with $F(c, u) = 1$, and building the tree in time $O(n)$ (this is easy to do if we assume that DLs are represented as intervals $[i, j]$, we omit the details).  Thus, the reconstruction phase does not take more time than the $B$ and $F$ phase. 
The total reconstruction time is therefore $O(pnV(\mathbb{T}))$.

\subsection*{Pseudo-code of IL\_ILC (and RF\_ILC) median: Algorithm~\ref{c2_dp_algorithm}}

\begin{algorithm}
    \textbf{Input:} structural trees $T_1, \dots, T_p$ over $[0, n+1]$, a boolean \textsf{input\_only}\\
    \textbf{Output:} An RNA tree $M$ minimizing 
    $\sum_{i=1}^p d(T_i,M)$. If \textsf{input\_only} is true, then $M$ may only contain leaf-sets from the input trees
    \begin{algorithmic}[1]
        \Function{IL\_Median}{$T_1,\dots,T_p$, \textsf{input\_only}}:
           \State $\hat{c}=\{\}$\Comment{// initializing DP table}
           \State optimum = {\tt optimal\_score}(0, n+1) \Comment{filling DP table}
           \State M = {\tt tree\_from\_leafsets}({\tt backtrace}(0, n+1))
           \State \Return M 
        \EndFunction
    \end{algorithmic}
    \begin{algorithmic}[1]
        \Function{\tt optimal\_score}{$i,j$}
        
        % //global variables: $c,T_1,\dots,T_p$,\textsf{input\_only}
            \If{$(i,j)$ in $c$} \Return $c[i,j]$
            \EndIf
            \State $\hat{c}[i,j] = +\infty$ 
            \State
            \State \emph{// Computing $\hat{c}_1[i,j]$}
            \For{$I$ in $\cup_{i=1}^p \mathcal{IL}(T_i)$ if $I$ contains $i$}
                \State $\text{score\_with\_}I=|\{i\mid I\notin \mathcal{IL}(T_i)\}|-|\{i\mid I\in\mathcal{IL}(T_i)\}|$
                \State $+\sum_{h\in\text{\tt holes}(I)} \text{\tt optimal\_score}(i_h,j_h)$
                \State $\hat{c}[i,j] = \min\left(\hat{c}[i,j],\text{score\_with\_I}\right)$
            \EndFor
        \State
        \If{not \textsf{ILC}}
        \State \emph{// Computing $\hat{c}_2[i,j]$}
        \For{$k\in[i+1,j]$}
            \State $\alpha = \text{\tt mwIS\_interval\_graph}(G_{k+1,j})$ 
            \State \hfill \emph{// line above requires $\hat{c}[u,v]$ for all $i<u<v<j$}
            \State $\hat{c}[i,j] = \min\left(\hat{c}[i,j],p+\hat{c}[i+1,k-1]-\alpha\right)$
        \EndFor
        \EndIf
        \State \Return $\hat{c}[i,j]$
        \EndFunction
    \end{algorithmic}
    \begin{algorithmic}[1]
        \Function{\tt backtrace}{$i,j$}
            \For{$I$ in $\cup_{i=1}^p \mathcal{IL}(T_i)$ if $I$ contains $i$}
            \State $\text{score\_with\_}I=|\{i\mid I\notin \mathcal{IL}(T_i)\}|$
            \State $+\sum_{h\in\text{\tt holes}(I)} c[i_h,j_h]$
            \If{$\text{score\_with\_}I=c[i,j]$}
            \State \Return $[I]+\text{\tt concatenate}(\{\text{\tt backtrace}(x,y)\mid (x,y)\in\text{\tt holes}(I)\})$
            \EndIf
            \EndFor
        \If{not \textsf{ILC}}
        \For{$k\in[i+1,j]$}
            \State $\alpha = \text{\tt mwIS\_interval\_graph}(G_{k+1,j})$ 
            \If{$\hat{c}[i,j]=p+\hat{c}[i+1,k-1]-\alpha$}
                \State \emph{// \text{\tt mwIS}: a maximum weight IS of $G_{k+1,j}$}
                \State \Return $[\{i,k\}\cup [k+1,j]\setminus \text{\tt mwIS} ]$
                \State $+\text{\tt backtrace}(i+1,k-1)$
                \State $+\text{\tt concatenate}\left(\{\text{\tt backtrace}(x,y)\mid x,y \in \text{\tt mwIS}\}\right)$
            \EndIf
        \EndFor
        \EndIf
        \EndFunction
    \end{algorithmic}
    \caption{Folding-like, dynamic programming algorithm for computing the
    IL Median of a set of input trees.}
    \label{c2_dp_algorithm}
\end{algorithm}

\subsection{Supplementary Figures}

\end{document}